\documentclass[preprint,12pt]{elsarticle}
\usepackage{inputenc}
\usepackage{verbatim}
\usepackage{psfrag}
\usepackage[english]{babel}	
\usepackage{amsmath,amsfonts,amsthm} 
\usepackage{amssymb,latexsym}
\usepackage{graphicx}	
\usepackage{newlfont}
\usepackage{psfrag}
\usepackage[a4paper,top=2.5cm,left=2.5cm,bottom=2.5cm,right=2cm]{geometry}
\usepackage{hyperref}
\usepackage[table]{xcolor}

\definecolor{mlib0}{HTML}{1f77b4}
\definecolor{mlib1}{HTML}{ff7f0e}
\definecolor{mlib2}{HTML}{2ca02c}
\definecolor{mlib3}{HTML}{d62728}
\definecolor{mlib4}{HTML}{9467bd}
\definecolor{mlab1}{HTML}{0072bd}
\definecolor{mlab2}{HTML}{a2142f}
\definecolor{mlab3}{HTML}{edb120}
\definecolor{mlab4}{HTML}{7e2f8e}

\newcommand{\mylab}[3]{\raisebox{#2}[0mm][0mm]%
           {\makebox[0mm][l]{\hspace*{#1}{#3}}}}

\newcommand{\fullsquare}{\mbox{$\blacksquare$}}

\newcommand{\fullcirc}{\mbox{{\Large$\bullet$}}}

\definecolor{dkgreen}{rgb}{0,0.6,0}
\definecolor{gray}{rgb}{0.5,0.5,0.5}
\definecolor{mauve}{rgb}{0.58,0,0.82}

\newcommand{\mH}{\mathrm{H}}
\newcommand{\deltah}{{\mathrm{\delta}h}}
\newcommand{\ett}{\ensuremath{T_\text{ett}}}
\numberwithin{equation}{section}	
\numberwithin{figure}{section}		
\numberwithin{table}{section}		

\def\mat#1{\mbox{\boldmath $\mathsf #1$}}
\newcommand{\change}[1]{\textcolor{black}{#1}}
\graphicspath{{./figures/}}

\journal{Journal of Computational Science}

\begin {document}
\begin{frontmatter}

\title{\change{A low-storage method consistent with second-order statistics for
time-resolved databases of turbulent} channel flow up to $Re_\tau=5300$}


\author{Alberto Vela-Mart\'in}
\author{Miguel P. Encinar}
\author{Adri\'an Garc\'ia-Guti\'errez}
\author{Javier Jim\'{e}nez}

\address{School of Aeronautics, Universidad Polit\'ecnica de Madrid}

\begin{abstract}
Wall-bounded flows play an important role in numerous common applications, and have been
intensively studied for over a century. However, the dynamics and structure of the
logarithmic and outer regions remain controversial to this date, and understanding their
dynamics is essential for the development of effective prediction and control strategies, and for the construction of a complete theory of wall-bounded flows. Recently, the use of time-resolved
direct numerical simulations of turbulent flows at high Reynolds numbers has proved useful
to study the physics of wall-bounded turbulence, but a proper analysis of the logarithmic
and outer layers requires simulations at high Reynolds numbers in large domains, making the
storage of complete time series challenging. 
In this paper a novel low-storage method for
time-resolved databases is presented. This approach reduces the storage cost of time-resolved databases by storing filtered flow fields that target the large and intermediate scales,
while retaining all the information needed to fully reconstruct the flow at the level of filtered flow fields and complete second-order statistics. 
\change{This is done by storing also the filtered turbulent stresses,
allowing to recover the exact effect of the small scales on the large and intermediate scales.}  
A significant speed-up of the computations is achieved, first, by relaxing the numerical resolution, which is shown to affect only the dynamics close to the wall, but not the large scales stored in the database, and, second, by exploiting the \change{computing power and efficiency} of GPU co-processors using a new high-resolution hybrid CUDA-MPI code.
This speed-up allows running for physically meaningful times to capture the dynamics of the large scales. 
The resulting temporally resolved large-scale database of a turbulent channel flow up to $Re_\tau=5300$, in large boxes for long times, is briefly introduced, showing significant indicators of large-scale dynamics with characteristic times of the order of up to eight eddy turnover times.

\end{abstract}


\end{frontmatter}

\section{Introduction\label{sec:intro}}

Wall-bounded flows are fundamental building blocks of many industrial
applications as well as natural phenomena, and developing accurate models for their prediction and control is a crucial
challenge for the next decades. One fourth of the energy in advanced economies is used in
transportation, and about 20\% of that amount is dissipated in wall-bounded turbulent flows.
Therefore, 5\% of the total energy, and a disproportionate amount of the resulting CO$_2$
emissions, are spent that way \cite{jimenez2013near}.
Wall turbulence is the main contributor to
aerodynamic friction, and is responsible for most of the pressure drop in internal flows
such as pipelines. Its importance cannot be overemphasised for the \change{demanding energetic challenge} of this
century. Among wall bounded flows, channels are the simplest to simulate numerically, i.e. they achieve the largest \textit{Re} for the same numerical complexity, and
represent a valuable tool for the accurate study of these phenomena. Their direct numerical
simulation (DNS) has been a basic component of their study \cite{kim1987turbulence,
mos:kim:man:99, ala:jim:03, hoy:jim:2006, leemoser15}.

While the behaviour of turbulence in the viscous layer near the wall has been reasonably
well understood for some time \cite{jim:moi:91, ham:kim:wal:95}, the dynamics of the flow farther
from the wall is less clear. One of the most vexing remaining problems is the logarithmic
layer that separates the near-wall and outer regions. It is in this layer that most of the
velocity drop of a boundary layer takes place at high Reynolds numbers, and where the
transition between the very different length-scales of turbulence in the two regions occur.
The logarithmic layer has been studied experimentally for a long time,
but numerical data have only been available in the last decade because it only exists at
relatively high Reynolds numbers. Although different limits have been
proposed \change{\cite{pop:00, jim:2012, leemoser15}}, it is safe to assume that the logarithmic layer extends from above $x_2u_\tau/\nu \approx 150$ to $x_2/h=0.15$,
where $x_2$ is the distance from the wall, $u_\tau$ is the friction velocity, $\nu$
is the kinematic viscosity, and $h$ is the boundary-layer thickness or the channel
half-width. Therefore, a logarithmic layer does not exist if the friction Reynolds number
is $Re_\tau=u_\tau h/\nu\lesssim 1000$, and it is only appreciably wide if $Re_\tau$ is
substantially higher.
The first DNSs of channel flows at $Re_\tau = 180$--$950$ had no
logarithmic layer in the sense just mentioned \cite{kim1987turbulence, mos:kim:man:99, ala:jim:03}.
Since then, $Re_\tau$ has increased steadily,
and simulations with a short logarithmic layer have become available. The current
state of the art is $Re_\tau\sim4000$--$8000$ \cite{ber:pir:orl:14,loz:jim:2014,leemoser15,
yamamoto2018numerical}, where the logarithmic layer extends over a scale ratio of 5 to 8 in the 
wall-normal direction.

Another open question in wall-bounded turbulence is the origin and dynamics of the large scales in the outer 
region, which extends above $x_2\sim0.1h$.
These scales play a relevant role in wall-bounded flows as they contain most of the total kinetic energy and Reynolds stresses of the flow,
especially at high Reynolds numbers \cite{jim:98,kim:adr:99,ala:jim:01,liu:adr:han:01,ala:jim:zan:mos:04},
but their study has been hindered by the difficulty to obtain quality data.
In the case of channels, the flow between two infinite parallel walls is modelled by imposing periodic boundary conditions 
in the two wall-parallel dimensions, where the period represents the size of the numerical box. 
It was soon realised that the size of the box is an important parameter 
that conditions the size of the largest `well-resolved' structures \cite{jim:moi:91,ala:jim:zan:mos:04,flo:jim:10,loz:jim:2014}. 
The computational box has to be large enough for the structures in the outer region to be unconstrained,
in the same way that the grid has to be fine enough to capture the smallest eddies and to
reproduce the correct dissipation. A \emph{de facto} standard for large-box simulations is
(length~$\times$~span)$=(8\pi \times 3\pi) h$, which is designed to resolve the length of the longest channel features $O(20h)$ \cite{jim:2012}. Experiments and simulations
in boxes up to $(60\pi \times 6\pi) h$ \cite{loz:jim:2014}, have shown that this is a
reasonable size to properly reproduce the dynamics of the largest structures that arise in channels.



The increasing availability of computational resources has provided DNSs of turbulent channel flow 
whose small and large scales are well resolved, but which focus mainly on producing high-quality snapshots and 
average statistical properties.
In general, these simulations do not offer {\it a posteriori} access to the dynamics of the flow and 
their practical utility is limited in this sense.
A much more valuable insight into the dynamics of these flows is possible by also storing temporally resolved data. 
While all DNSs are by definition temporally resolved while they are being computed, the possibility of
storing and post-processing temporally resolved time series, instead of a few
independent snapshots, has become possible only recently.
Since the early efforts in the \change{1990s}
\cite{rob:91}, the use of temporal series at higher Reynolds
numbers has become a relevant tool of turbulent research, especially in wall-bounded flows \cite{flo:jim:10,loz:jim:2014}.
These temporal series, which are freely available to the community \cite{website:dmt, summerprogram}, mark an inflection point in turbulence research,
because they allow several groups to test hypotheses on the same
data, and to refine them interactively. The limiting resources to produce useful databases are storage space and post-processing time. The bottleneck of these simulations lies on storing and analysing temporally and spatially
resolved time series of high-Reynolds numbers channels in large boxes,
specially when the dynamics of the large scales are targeted.
Part of the reason for the high cost of generating and storing useful temporal series of the
large scales is that the simulation time has to be long enough to capture their
relatively slow temporal dynamics. In the logarithmic layer, the lifetime of structures centred at a distance $x_2$ from 
the wall is $u_\tau T\approx 6 x_2$, or about an eddy-turnover time ($\ett=h/u_\tau$) for structures
at $x_2/h=0.15$ \cite{flo:jim:10}.
Reasonable statistics require that the simulation time is $u_\tau t/h\gg 1$.

 
In the previous paragraphs, we have stressed the requirements of a turbulent database directed
towards the study of the large scales of a turbulent channel flow: high Reynolds numbers, 
an adequate box size, and storing temporally resolved and sufficiently long series.
Reviewing the characteristics of the currently available databases,
it becomes apparent that none of them meets all these desirable properties, although some
combinations of two of them can be found. For example, recent non-time-resolved simulations
at high Reynolds numbers in large boxes can be found in \cite{leemoser15}. Time-resolved series
at moderate Reynolds numbers, $Re_\tau=1000\mbox{--}2000$, have been available 
for some time \cite{flo:jim:10,loz:jim:2014}, but in small box sizes. 
A recent addition to the available databases is \cite{pha_jhu}, which includes an $Re_\tau=1000$
time-resolved channel in a large box, although only for a relatively short temporal period $(u_\tau
t/h\approx 1.4)$. While large boxes require larger computational resources than smaller ones
at the same Reynolds number, both share the limiting factor of storing, sharing, and
post-processing the computed data. For example, assuming a maximum data storage of 200 TB,
and a target $Re_\tau =4000$, the choice is between saving 500 snapshots of a large
$(8\pi \times 3\pi) h$ box, or 6500 snapshots of a simulation in a smaller $(2\pi \times \pi) h$ box.
The latter database would provide six \ett{} with a reasonable time interval between snapshots
\cite{lozano-time}, while the former would only contain 0.5\ett.

In this paper we present a novel method to produce affordable, high quality, time-resolved databases of
turbulent channel flow aimed at the study of the intermediate and large scales.
This method, which can be easily extended to other flows, relies on two aspects.
First, a dynamically meaningful reduction of the amount of information stored,
which covers only large scales while retaining the effect of the small scales,
and, second, the speed up of the computations by means of a controlled reduction  of the numerical resolution,
which only affects the small scales, but not the scales of interest stored in the database.
Further speed-up is achieved by implementing a new parallel GPU solver, 
which exploits the advantages of powerful heterogeneous architectures.
This paper covers the computation and generation of a database for a state-of-the-art
Reynolds number, $Re_\tau \sim 5000$, in a $(8\pi \times 3\pi) h$ box, providing time-resolved
data for the unprecedentedly long time of 36\ett.

The storage limitations can be alleviated if one mostly wishes to study the dynamics of the
large and intermediate motions, which, as mentioned above, are the ones that require large
boxes. Although DNS simulations have to be properly resolved to reproduce the
physics of turbulence, large and small scales are relatively independent, and do not need to be
post-processed together. Recent simulations have shown that damping \cite{hwa:cos:11}, or
even removing \cite{miz:jim:2013,dong17} the small scales near the wall, has negligible
effects on the logarithmic-layer structures, and it has been known for some time that
large-eddy simulations, which have no small scales, reproduce the large structures correctly
\cite{scov2001}. Supported on these evidences,
we store only accurate data obtained by \emph{a posteriori} filtering of the DNS at \change{a prescribed scale.}
\change{We store snapshots with enough temporal resolution (in the 'Nyquist' sense) to capture the dynamics of at the scale of the filter.
This allows to reproduce the temporal dynamics of any scale above the filter scale, while alleviating the storage requirements by reducing both the spatial and temporal resolution of the database.}
The dynamics of the small scales are only preserved in the form of
on-the-fly post-processed statistics.
The novelty of our approach lies in retaining also their dynamical effect on the filtered data 
by storing also the filtered Reynolds stresses.
This approach allows a complete reconstruction of the dynamics of 
the three velocity components above the database filter.
Channel simulations in a large box \change{become} feasible, and \change{open} new possibilities for studying the
temporal evolution of accurately computed large-scale motions in wall-bounded flows. 
Discarding the dynamics of the small scales is justified on the availability of time-resolved
simulations in boxes which, although smaller, are large enough to allow the study of these scales \cite{flo:jim:10,loz:jim:2014,pha_jhu}.

Since our aim is to capture the dynamics of the intermediate and large scales only,
we require that the effect of the dissipative scales on these scales is well reproduced,
but not that their dynamics are faithfully resolved to the level of high-quality DNSs.
Considering that a relevant fraction of the total computational resources in DNSs 
is used to resolve the dissipative scales, we obtain a substantial speed-up in our simulations
by adequately relaxing the standard resolution requirements, while ensuring that the dynamics of the large
scales are not affected. 
This approach has been widely used in homogeneous isotropic turbulence,
for which simulations at high Reynolds numbers are produced by reducing the resolution of the small scales \citep{ishihara2009study}. 
In this simulations, the average locality of scale interactions ensures that the effect of the insufficient resolution is 
confined to the small scales, but does not affect the inertial scales.
This approach not only reduces the number of grid points, but also relaxes the stability conditions of the temporal 
integrator, providing a significant speed-up in terms of computational time per \ett.  
We reduce the numerical resolution in the wall-parallel directions to reduce the computational requirements,
and modify the distribution of points close to the wall to alleviate the stability requirements of the temporal integrator.
We compensate this reduction of the numerical resolution by implementing spectral methods in the wall-parallel directions,
and high-order, high-resolution compact finite differences in the wall normal direction. 
These modifications of the standard DNS requirements allow us to run for unprecedented long times,
and do not affect the truncated data stored in the database.
We present results that indicate that a reduction of the numerical resolution with respect to
standard high-quality DNSs is acceptable for simulations that target the large scales of wall-bounded flows.

In this work we have also validated the suitability of GPUs for large-scale direct numerical
simulations of turbulent flows using high resolution and spectral methods. 
Although GPUs have been part of many supercomputers for some
time, their use for solving fluid mechanics problems is not general. Some previous codes have
made use of many GPUs to gain a considerable speed-up \cite{kha:per:13,kar:pol:eka:14,zhu2018afid}, but
they generally use low-resolution spatial discretization schemes that make them less
attractive for high-Reynolds number turbulence. Our results prove that the use of GPUs can also be
advantageous for the computation of communication-intensive simulations such as those needed
to compute turbulence with spectral and high-order compact schemes.
For implementations that spend about 40\% of
the time transferring data, asynchronous overlapping between communications and computation
can speed up simulations by a factor of two with respect to a standard CPU code running on
the same platform. For large simulations spanning months, this advantage is crucial. As
network bandwidth and clusters architecture improve, and the fraction of time spent
communicating decreases, the advantage of using GPUs should be even clearer. When the next
step in supercomputation is taken, and new exascale machines become available, efficient
devices will be required to keep energy consumption down to a reasonable level. GPUs have
proved to be an efficient and fast solution, and are present in ever more supercomputing centres, as they
deliver good performance while keeping energy consumption low. The code used in this
investigation takes advantage of the use of GPUs on a large scale, and represents a 
step into using future heterogeneous CPU/GPU exascale architectures for DNS simulations.
The code is publicly available \cite{website:code}. 

This paper is organised as follows. The algorithm and code employed are
briefly described in 
\S \ref{sec:methods}.
Section \ref{database} describes the storage approach, the simulations
performed for the database, the main statistics of those simulations, and their validation.
Finally, conclusions are offered in \S \ref{sec:conclusions}


\section{Methods \label{sec:methods}}

\subsection{Simulation algorithm}

\begin{figure}[t]
\centering
\includegraphics[width=0.8\textwidth]{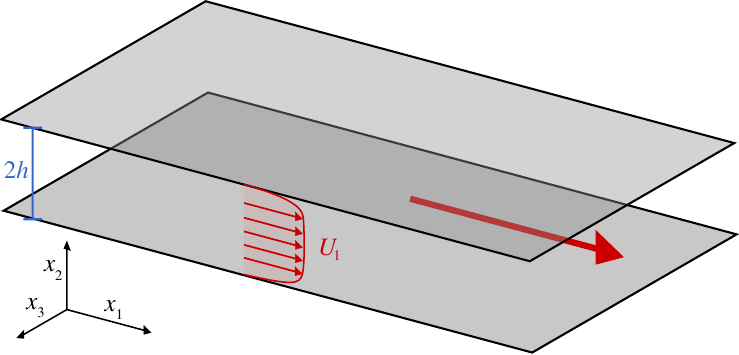}\\[20pt]

\includegraphics[width=\textwidth]{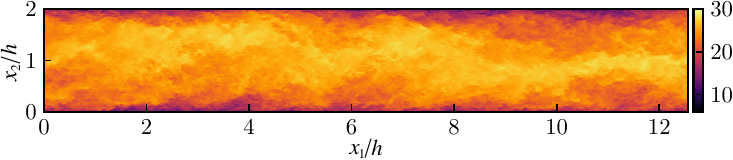}
\label{fig:1}
\caption{Top: Sketch of channel flow between two parallel walls. Bottom: snapshot of $u_1^+$ in the F5300 simulation (see table \ref{tab:2})}
\end{figure}

We consider the incompressible turbulent flow between two parallel planes separated \change{by} a distance of $2h$. Both the streamwise, $x_1$, and the
spanwise, $x_3$, directions are periodic, with period $L_1$ and
$L_3$, respectively. A constant mass flux is imposed in the streamwise direction by a
time-variable mean pressure gradient. Due to incompressibility and impermeability at the wall, 
the averaged velocity in the wall-normal direction at a given height, $x_2$, is zero at all distances from the wall.
The mean value of the spanwise velocity component is allowed to move freely, since no pressure gradient is applied along that direction, 
but is non-drifting in the long-term average. A sketch of the flow is shown in figure \ref{fig:1}.
Magnitudes expressed in `wall' units, which are constructed from the friction velocity $u_\tau$ and the
kinematic viscosity $\nu$, are denoted by a `+' superindex, so that $Re_\tau=h^+$. Upper
case symbols are used for ensemble-average quantities, such as the mean velocity profile
$U_1$, while primes are reserved for root-mean-square intensities.

%

The evolution equations for the fluid velocity vector, $\boldsymbol u=[u_i]$, are the Navier--Stokes equations for an
incompressible fluid,
\begin{equation}
\partial_t u_i = -u_j\partial_j u_i -\partial_i p + \nu \partial_{kk} u_i + \delta_{1i}f,
\label{eq:e1}
\end{equation}
\begin{equation}
\partial_i u_i = 0,
\label{eq:e2}
\end{equation}
where $u_1$, $u_2$, and $u_3$ are the streamwise, wall-normal and spanwise velocities,
$p$ is the kinematic pressure, $\delta_{ij}$ is the Kronecker delta, and $f$ the time-variable mean pressure gradient imposed to keep a fixed mass flux.
Repeated indices imply summation.
For convenience,
equations (\ref{eq:e1}) and (\ref{eq:e2}) are transformed into an \change{equation} for the vorticity in the wall-normal direction, ${\omega_2}=\partial_3 u_1 - \partial_1 u_3$, an equation for the Laplacian of the wall-normal velocity,
$\partial_{kk} u_2$, and two equations for the mean velocity profiles, $U_1$ and $U_3$, thus removing the computation of the pressure \cite{kim1987turbulence}.
The reader is referred to \ref{sec:alg} for a brief review of this method.

The equations are projected on a Fourier basis with $N_1/2$ and $N_3/2$ modes in each periodic direction. In real space, the mesh is uniformly spaced,
with  $N_1$ and $N_3$ collocation nodes chosen to provide the desired numerical resolution.
Non-linear terms are computed using a pseudo-spectral method and are fully dealiased by zero padding the Fourier fields to $3N_1/4$ and $3N_3/4$ modes \cite{can:hus:qua:zan:88}. In the wall-normal direction, the equations are discretized on a
non-uniform mesh of $N_2$ points, adjusted to keep, at all wall distances, an approximately constant resolution in
terms of the local Kolmogorov scale \cite{flo:jim:06}.
Derivatives in this direction are
computed directly on the mesh using seven-point compact finite differences 
with spectra-like resolution \cite{lel:92,gamet1999compact}. 
\change{These} numerical approximations have very high resolution properties and high order, but we find that they may become numerically unstable close to the wall, especially at high Reynolds numbers, if the mesh is not correctly devised.  
We find numerical instabilities in the implicit viscous step of 
the Laplacian of the wall-normal velocity, which we resolve by optimally 
modifying the physical mesh close to the wall. Details of the method are provided in \ref{sec:enf}.
A semi-implicit third-order low-storage Runge--Kutta is used for
temporal integration \cite{spalart1991spectral}.

\subsection{Code}

We implement a novel hybrid CUDA--MPI code to run the simulations on many distributed graphic processing units (GPUs). All numerical algorithms use exclusively the GPU,
and the CPU is used only for node-to-node communications. 
The code relies on the highly optimised CUFFT library \cite{website:cufft} to perform Fourier transforms. Custom CUDA \change{kernels} have been devised for the rest of algorithms, including the compact-finite differences.

A plane-plane decomposition is used to partition the computational domain among GPUs.
A configuration in $x_2x_3$-planes is used to provide local access to full wall-normal pencils, which simplifies the inversion of banded matrices required by the compact finite differences.
For the computation of the non-linear terms, the domain is reorganised in $x_1x_3$-planes to perform inverse and direct Fourier transforms in wall-parallel planes.
This decomposition allows to implement the zero-padding dealiasing locally in each GPU, reducing the amount of communications required to compute the non-linear terms.
All the local transpose operations required for the global transposes are performed in the GPU to leverage their high memory
bandwidth. Several strategies are used for the MPI communication part, from all-to-all to send/receive. 
GPU execution and communications are performed asynchronously, 
and particular care is taken to overlap as much as possible communications and computations. A detailed description of the optimisation procedure is presented in \ref{sec:opt}.

The code shows good scaling in up to 1024 GPUs, proving the suitability of GPUs for the computation of
DNSs at high Reynolds number using high-resolution spectral methods.
We also report a strong dependence of the scaling and performance on the balance between computation power, and network bandwidth. Details of the scaling and performance of the code are compiled in \ref{sec:scal}.

\section{A second-order consistent, low-storage time-resolved database\label{database}}

\subsection{Methodology}
A time-resolved turbulent channel flow simulation at $Re_\tau\sim5000$ spanning 36\ett{} in a standard large box ($L_1=8\pi h$ and $L_3=3\pi h$)
would produce 55 PB of data if the evolution of all the flow scales were to be retained
\cite{lozano-time, pha_jhu}. 
Most of the
information contained in that disproportionately large amount of data is used to represent
the dynamics of viscous scales, which can be studied in much more affordable simulations in
smaller boxes.

To alleviate the storage requirements of the time-resolved database,
we only store the inertial and large scales above the dissipative range, reducing by orders of magnitude the size of the database. 
This reduction comes from the lower spatial and temporal resolution
necessary to accurately capture the dynamics of the large scales. 
The inertial range starts at approximately $50\eta$, where $\eta$ is the Kolmogorov length-scale
\cite{kol:41}), which translates roughly into $100\nu/u_\tau$ \cite{jim:2012}.
If we aim to time-track an eddy of size $l_e$ advected at velocity $u_e$ with respect to the mean velocity profile, the characteristic time between snapshots, $\Delta t_e$, needed to accurately represent its motion is of the order of $l_e/u_e$.
We target eddies of size larger than $l_e^+\sim100$, and assume that they are advected with respect to the mean profile at approximately the root mean square of the velocity perturbations.
The most restrictive condition arises from the stream-wise velocity perturbations in the logarithmic layer, where $u_1'^+\sim 2$.
Thus, a reasonable estimate is $\Delta t_e^+ \approx50$, which implies storing around $Re_\tau/\Delta t_e^+\sim100$ snapshots per \ett.

\begin{figure}
    \centering
    \includegraphics[width=\textwidth]{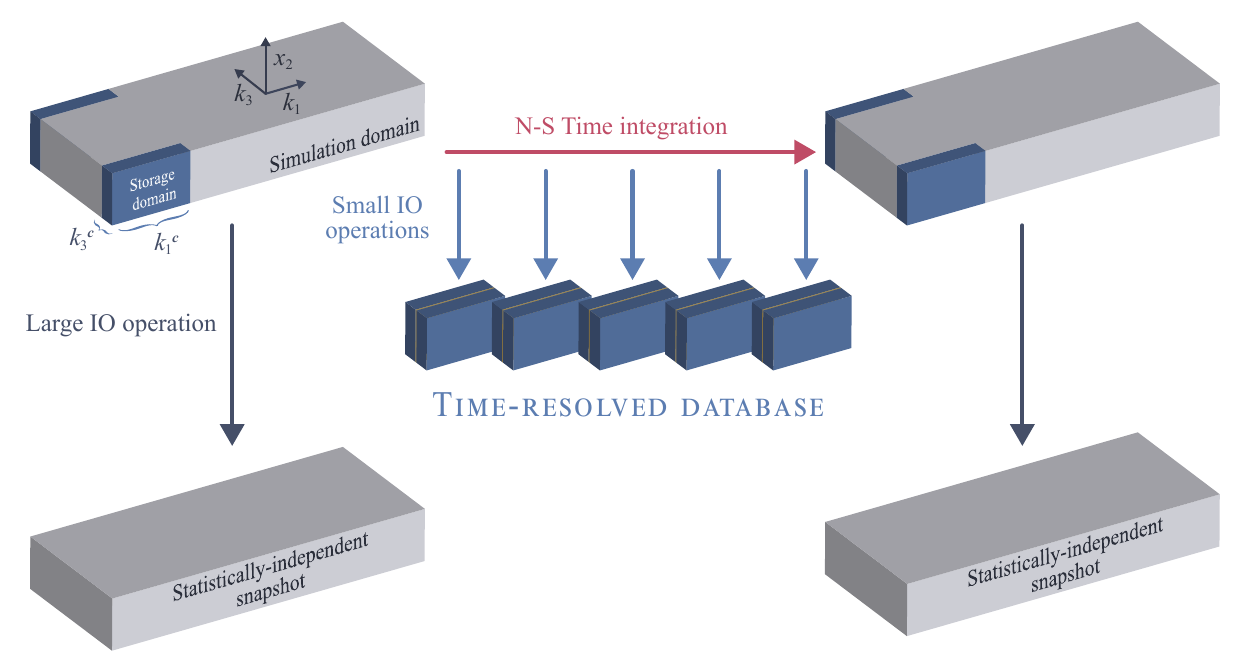}
    \caption{Schematic representation of the generation process of the low-storage second-order consistent database. The grey blocks represent the full field used to run the simulation, and which are stored every $0.5T_{ett}$, and the blue blocks the Fourier-truncated fields that are stored in the time-resolved database.}
    \label{fig:mono_storage}
\end{figure}

\begin{figure}[t]
    \centering
    \includegraphics[width=.8\textwidth]{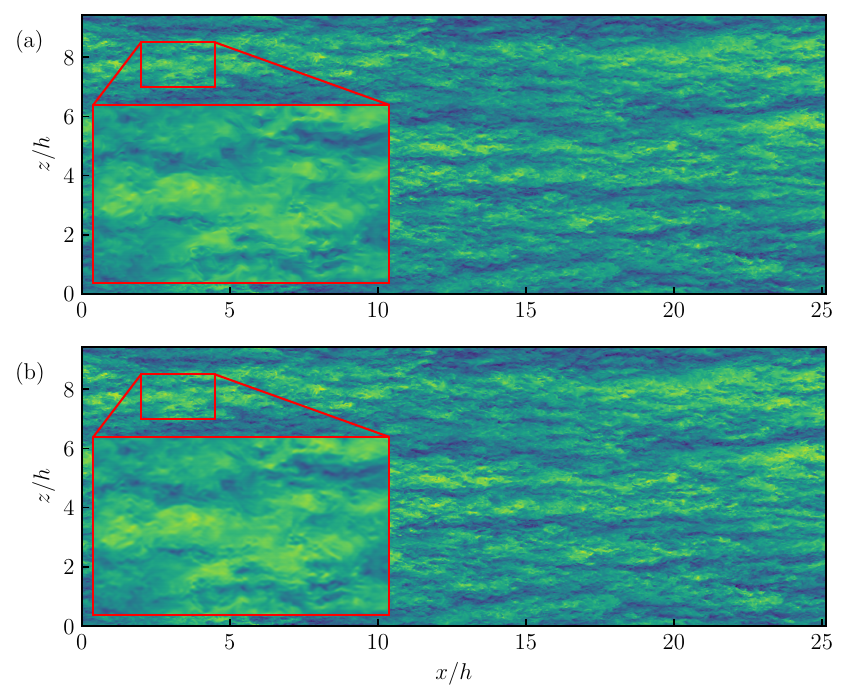}
    \caption{Sections of the streamwise velocity at $Re_\tau=2000$ and distance from the wall $x^+_2=150$. (a) DNS resolution ($\Delta x^+ \approx 12, \Delta z^+ \approx 9$). (b) New database resolution ($\Delta x^+ \approx 100, \Delta z^+ \approx 37$)(a). The inset magnifies a region four times. Data from \citep{hoy:jim:2006}.}
    \label{fig:planos_2000}
\end{figure}

To remove the small scales from the database, we truncate the velocity fields in Fourier space before storing them. This operation is similar to applying a sharp cut-off Fourier filter in the wall parallel directions,
\begin{equation}
\begin{aligned}
\widehat G(k_1,k_2,x_2)&=1 \text{ if } k_1<\change{k^c_1}  \text{ and }  k_3<k^c_3,  \\ 
\widehat G(k_1,k_2,x_2)&=0 \text{ otherwise}, 
\end{aligned}
\label{fou_fil}
\end{equation}
where the cut-off wavenumbers, $k^c_1$ and $k^c_3$, determine the smallest scales retained in the database, $l^c_1=\pi/k^c_1$ and $l^c_3=\pi/k^c_3$. Filtered quantities are marked with $(\overline{\cdot})$.
We do not perform any mode truncation or filtering in the wall-normal direction.
In figure \ref{fig:mono_storage}, we show a schematic representation of the methodology employed to produce the database.   
We retain scales above ${l^c_{1}}^+=120$ in $x_1$, and ${l^c_{3}}^+=60$ in $x_3$, and store the fields with a temporal resolution of $\Delta t_e^+=50$.
In figure (\ref{fig:planos_2000}), we show two streamwise velocity fields at $x_2^+$ for $Re_\tau\approx2000$, with and without Fourier truncation. 
The details of the large scales are indistinguishable to the naked eye.
The difference between the characteristic lengths in $x_1$ and $x_3$ accounts for the elongated
nature of the velocity perturbations in channel flows \cite{hoy:jim:2006}.
For $Re_\tau\approx5300$, this truncation reduces the number of collocation points in $x_1$ and $x_3$ \change{from} $N_1=6144$ to $M_1=L_1 Re_\tau/l_{1}^+\approx 1000$ and from $N_3=4196$ to $M_3=L_3 Re_\tau/l_{3}^+\approx 800$, reducing the size of each field by a factor of approximately $30$. In addition to the time-resolved low-storage database, we also store complete velocity fields roughly every $0.5T_{ett}$.

The effect of the removed small scales into the large scales is not negligible, and
impossible to recover from the filtered velocity field. To partially overcome \change{these}
limitations, we also store the truncated Reynolds-stress tensor $\overline{u_iu_j}$, which are the only
second-order non-linear quantities that directly appear in the equations of motion of the truncated velocity fields.
From the six truncated components of this tensor, 
it is possible to fully reconstruct the effect of small-scale dynamics on the filtered
velocities, preserving all the dynamical information of the evolution of the large scales. 
Finally, we also stored the filtered enstrophy, which is a relevant quantity in turbulence dynamics, and cannot be recovered from the filtered stress tensor.
The ten variables of the database are $\overline{u_i},\, \overline{u_i u_j}$, and 
$\overline{\omega_i \omega_i}$. 
The filter operation commutes with any homogeneous linear operator, allowing to perform exact \emph{a posteriori} computations of quantities of interest not stored in the database.
Some of the quantities that can be trivially computed from the database are the filtered
vorticities, $\overline{\omega_i} = \varepsilon_{ijk}\partial_j\overline{u_k}$, where
$\varepsilon_{ijk}$ is the fully antisymmetric Levi-Civita symbol, or the filtered kinetic energy,
$\overline{E} = \frac{1}{2} \overline{u_i u_i}$.
The computation of the filtered pressure is less
evident but equally possible. 
Taking the divergence of the momentum equation yields,
\begin{equation}
  \partial_{j}u_i\partial_{i}u_j = -\partial_{kk}p,
  \label{eq:1}
\end{equation}
where continuity, $\partial_i u_i = 0$, makes every other term vanish. This is the well-known Poisson equation for the pressure \cite{kim1989structure},
\begin{equation}
  \partial_{kk} \overline{p} = -\partial_{ij}(\overline{u_i u_j}),
  \label{eq:2}
\end{equation}
which involves only the cross-derivatives of the filtered Reynolds stress tensor, included in the database. An identity worth noting is,
\begin{equation}
\partial_{ij}(\overline{u_i u_j}) = \overline{S_{ij}S_{ij}} - \frac{1}{4}\overline{\omega_i \omega_i},
\end{equation} 
where $S_{ij}=\tfrac{1}{2}(\partial_i u_j + \partial_j u_i)$ is the rate-of-strain tensor.
This identity gives a recipe to compute the filtered magnitude of the rate-of-strain tensor from the filtered enstrophy and the filtered Reynolds stresses.

Finally, we stress the relevance of a closed second-order database for \emph{a priori} testing of LES models at high Reynolds numbers. The subgrid-scale \change{(SGS)} stress tensor can be exactly computed from the filtered velocities as,
\begin{equation}
  \tau_{ij} = \overline{u_iu_j} - \overline{u}_j\overline{u}_j,
\end{equation}
providing an full reconstruction in space and time of the effect of the sub-grid dynamics on the database. These data allows the exact testing of LES models directly against DNS data.

\subsection{Simulations}

\begin{table}[!t]
\begin{center}
\begin{tabular}{c c c c c c c c c} 
 \hline
  Name & $Re_\tau$ & $L_1/h$ & $L_3/h$ & $N_1\times N_2\times N_3$ & $M_1\times M_2\times M_3$ & $\ett$ & $N_\text{snap}$\\ 
 \hline
  F2000 & $2000$ & $8\pi$ & $3\pi$ & $2048\times512\times2048$ & $512\times512\times512$ & $14$ & $1300$\\ 
  F5300 & $5303$ & $8\pi$ & $3\pi$ & $6144\times1024\times4096$ & $1024\times1024\times1024$ & $36$ & $3900$\\
  L2000 & $2003$ & $8\pi$ & $3\pi$ & $4096\times633\times3072$ & -- & $10$ & $232$\\ 
  L5200 & $5200$ & $8\pi$ & $3\pi$ & $10240\times1536\times7680$ & -- & $7.8$ & --\\
 \hline
\end{tabular}
\caption{ New simulations (F2000 and F5300) and simulations used to validate them (L2000, from \cite{hoy:jim:2006} and L5200, from \cite{leemoser15}). 
$M_1\times M_2 \times M_3$ is the size of the truncated fields in the database, $N_d$ is the simulation resolution while $M_d$ is the storing resolution, \ett{} is the simulation time in eddy turnovers, and $N_\text{snap}$ is the number of snapshots available in the database.\label{tab:2}}
\end{center}
\end{table}

Table \ref{tab:2} summarises the characteristics of the simulations performed, as well as of those used for validation. 
The new simulations ran in PizDaint, a GPU-based supercomputer part of the PRACE Tier-0 network,  for a total of $65,000$ node hours on $128$
GPUs for F2000 and $1,400,000$ node hours on $512$ GPUs for F5300. The total size of the filtered
databases is $5.6$ TB and $100$ TB respectively. To ensure equilibrium before collecting the data,
the first 5 eddy turnovers run immediately after interpolating the initial condition were discarded. The initial
conditions were interpolated from similar Reynolds number simulations in equilibrium.

Since the database targets the large scales of the flow, the numerical resolution of the simulations  
is not so critical as long as the dynamics of the stored scales are unaffected. We will show that it is possible to slightly degrade the standard DNS resolution in the wall-normal directions from $\Delta x_1^+\approx12$ and $\Delta x_3^+\approx9$ to $\Delta x_1^+\approx24$ and $\Delta x_3^+\approx15$, while producing a relative error of less than 2\% in the retained scales. This allows in turn to run for approximately four times more eddy turnovers with the same computational time.

\begin{figure}
\raggedleft
\def\sca{0.86}
\includegraphics[scale=\sca]{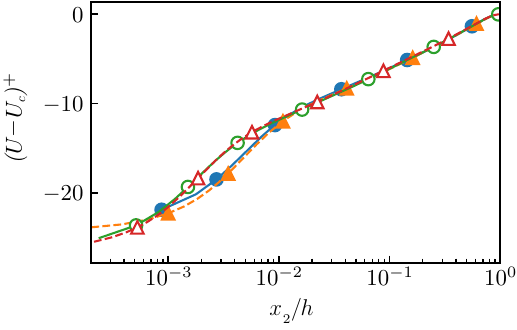}\mylab{-7.5cm}{4.45cm}{(a)}\hspace{0.78cm}
\includegraphics[scale=\sca]{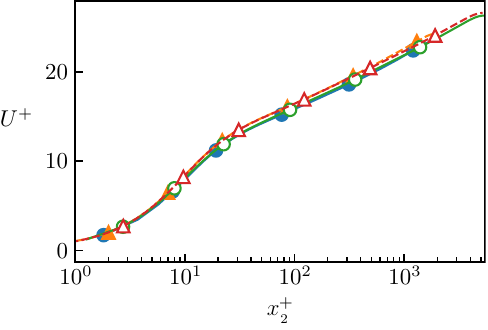}\mylab{-7.5cm}{4.45cm}{(b)}\hspace{0.14cm}\phantom{.}
\vspace{10pt}

\includegraphics[scale=\sca]{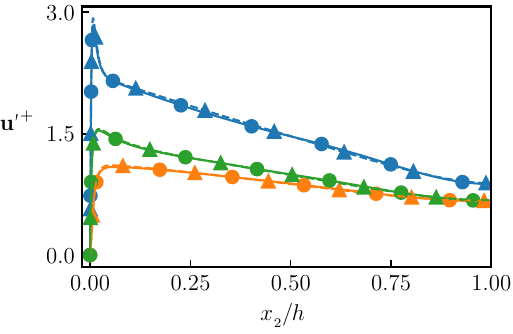}\mylab{-7.5cm}{4.45cm}{(c)}\hspace{0.65cm}
\includegraphics[scale=\sca]{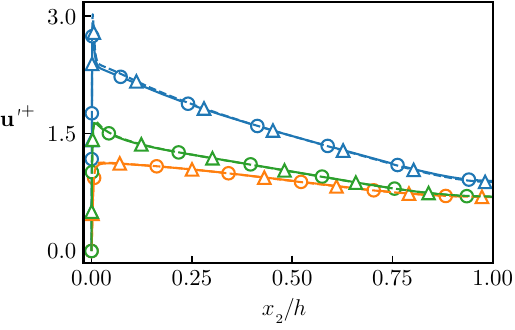}\mylab{-7.5cm}{4.45cm}{(d)}
\vspace{10pt}

\includegraphics[scale=\sca]{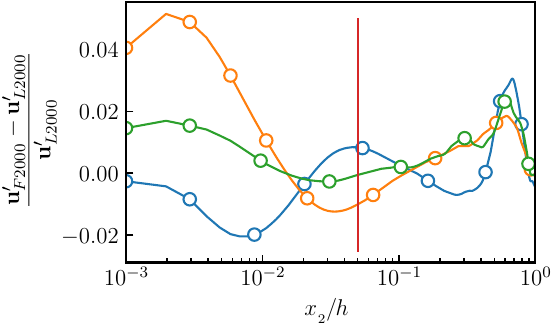}\mylab{-7.5cm}{4.45cm}{(e)}\hfill%
\includegraphics[scale=\sca]{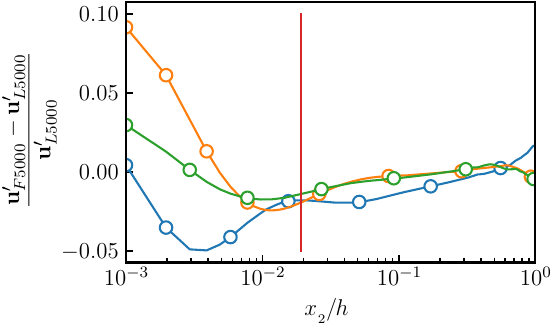}\mylab{-7.5cm}{4.45cm}{(f)}\hspace{0.12cm}\phantom{.}

\caption{ Mean profiles and root mean square of the perturbation velocities for the simulations in table \ref{tab:2}. \change{(a-d)} Symbols are triangles for the reference simulations and circles for the new ones, filled symbols are $Re_\tau \approx 2000$ and open ones $Re_\tau \approx 5000$. \change{(c-f) Colors are blue, $u_0$; orange, $u_1$; green, $u_2$}. (a) Mean velocity profile in velocity defect form for all the simulations. (b) Mean velocity profile for all the simulations. (b) Perturbation velocities rms for simulations at $Re_\tau \approx 2000$. (c) Perturbation velocities rms for simulations at $Re_\tau \approx 5000$. (d) Mean velocity profile plotted against inner units for all the simulations. (e) Relative error in the rms of the velocity perturbations for simulations at $Re_\tau \approx 2000$ (f) Relative error in the rms of the velocity perturbations for simulations at $Re_\tau \approx 5000$. The vertical line in \emph{(e,f)} is $x_2^+ = 150$.
\label{fig:stat}}
\end{figure}

\subsection{One-point statistics and energy spectra}

Figure \ref{fig:stat}(a-d) shows the main one-point statistics of the simulations described in table \ref{tab:2}. The mean profiles agree well with the ones of the full resolution simulations across all the channel, collapsing under the defect velocity form. When plotted in terms of the absolute velocity, they agree initially but depart later as a consequence of a constant shift in the value of the intercept constant, $C^+$ in the logarithmic law of the wall,
\begin{equation}
  \label{eq:lawofthewall}
  U^+_1 = \frac{1}{\kappa}\log(x_2^+) + C^+,
\end{equation}
where $\kappa$ is the K\'arm\'an constant.
The values of $\kappa$ and $C^+$ in the different simulations are computed by fitting \eqref{eq:lawofthewall} in the region $150 \nu/u_\tau < x_2 < 0.16 h$ for $Re_\tau \approx 2000$ simulations and $350 \nu/u_\tau < x_2 < 0.16 h$ for $Re_\tau \approx 5000$ simulations.
The difference in the value of the intercept constant is similar to the effect of transitional roughness, where a roughness function accounts for the departure of the mean profile due to the wall roughness \citep{jim:rou:04}. The profile of F2000 is shifted respect to L2000 by $\Delta C^+ \approx -0.12$,
and F5000 is shifted respect L5000 by $\Delta C^+ \approx -0.45$. 
In both cases the intercept constants are very close to the hydrodynamically smooth regime.
The root mean square (rms) of the perturbation velocities are plotted against their references in figures \ref{fig:stat}(c, d), and their relative errors, $(u_i^\prime - (u_i^\prime)_{\text{ref}})/(u_i^\prime)_{\text{ref}}$ are represented in figure \ref{fig:stat}(e, f). 
They agree well except near the wall.
Whereas the value of $u'_1$ is close to the reference simulations near the wall, 
$u'_2$ and $u'_3$ have more energy than the references, and their relative error is maximum at this height. Moving away from the wall,
the cross components converge to the reference value, and $u'_1$ lowers in intensity until the production peak of the buffer layer at $x_2^+ \approx 14$, where the relative error of the streamwise component is maximum. Farther away from the wall, at $x_2^+ > 100$, the deviation of either component is less than $2\%$ (or within the statistical error for the F2000 case). From the large error in the cross-velocity intensities near the wall, 
it is possible to trace the lower resolution of the grid as responsible for the difference in the intercept constant of the mean profile. Close to the wall, very small quasi-streamwise vortices populate the flow field, whose size is close to the resolution of the mesh in the reference simulations.
These vortices are not well resolved in the new simulations, which have lower resolution, as reflected by the error in the rms of the cross-velocity components. This lack of resolution affects the friction at the wall, and the intercept constant changes slightly as if it were under the effect of very fine roughness. However, this phenomenon does not affect the flow above the buffer layer, where the mean velocity profiles collapse under the velocity-defect form with the reference simulations. So do the intensity of the fluctuation velocities above $x_2^+ \approx 100$.


\begin{figure}
\centering
\def\sca{0.9}
\includegraphics[scale=\sca]{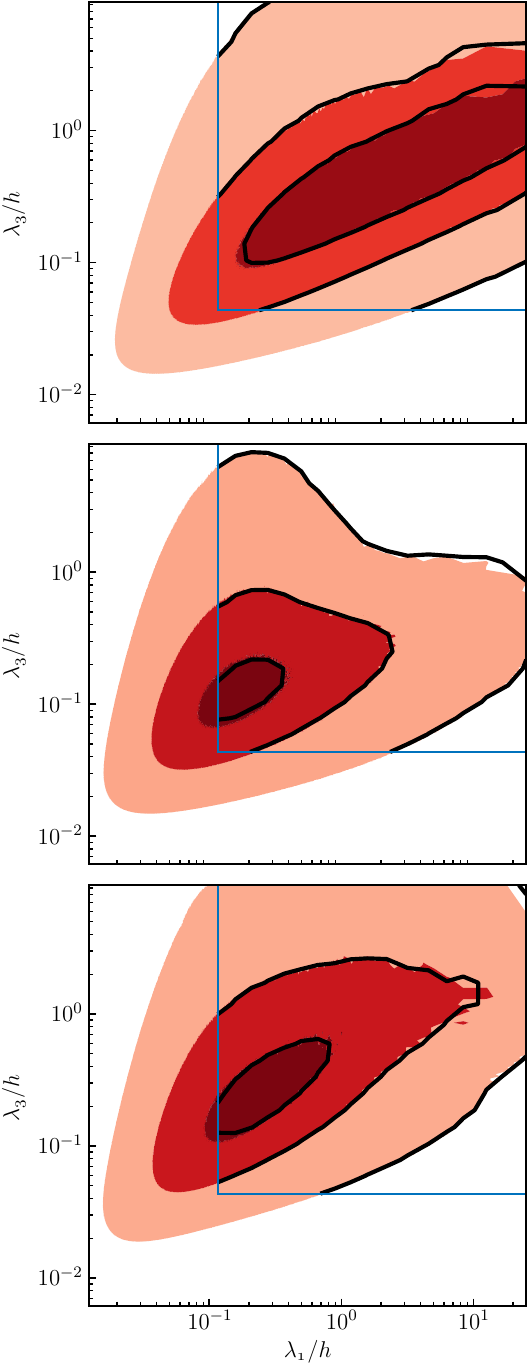}\mylab{-8.1cm}{20cm}{(a)}\mylab{-8.1cm}{13.3cm}{(b)}\mylab{-8.1cm}{6.5cm}{(c)}\hfill
\includegraphics[scale=\sca]{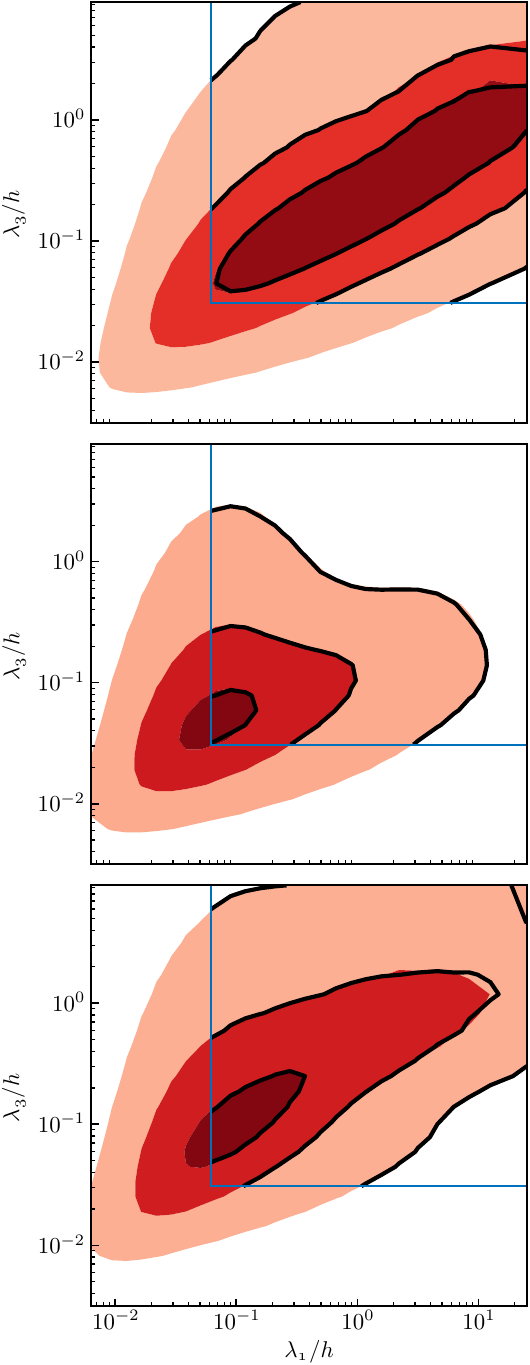}\mylab{-8.1cm}{20cm}{(d)}\mylab{-8.1cm}{13.3cm}{(e)}\mylab{-8.1cm}{6.5cm}{(f)}
\caption{Spectra of the velocity components at $x_2^+ = 150$ for the simulations in table \ref{tab:2}. (a-c) Spectra of $u_i$ for L2000 (red shaded patches) and F2000 (black line). (d-f) Spectra of $u_i$ for L5200 (red shaded patches) and F5300 (black line). The blue square marks the database resolution in all cases. Contours contain $90$, $50$ and $10\%$ of the spectral mass of the reference simulations, and the same contour level is used for the new ones. \label{fig:spec}}
\end{figure}

The same level of agreement is found in the two-dimensional spectra of the velocity components, defined as
\begin{equation}
E_{ii}(k_1, x_2, k_3) = \left\langle \widehat{u}_i(k_1, x_2, k_3)\widehat{u}_i^*(k_1, x_2, k_3)\right\rangle,\label{eq:spe}
\end{equation}
where the caret denotes the Fourier transform on the \change{wall-parallel} directions, $k_i$ are the spatial wavenumbers and $\lambda_i$ their corresponding wavelengths, the asterisk denotes complex conjugation and $\langle\cdot\rangle$ is the ensemble average. Figures \ref{fig:spec}(a-f) gathers the two-dimensional spectra of the three velocity components at $x_2^+ \approx 150$, which is the wall-normal location closest to the wall within the logarithmic region, where the reduced numerical resolution of the simulation should be most critical for the database.
The spectra of the new simulations agree remarkably well with the references over all the scales retained by the cutoff filter. As expected, the wall-normal velocity component is the worst resolved, with $\approx 13\%$ of its energy below the filter cutoff for F5300. The streamwise and spanwise velocities retain more than $98\%$ and $93\%$ of their energy respectively. As with the one point statistics, we have no indication that the reduced resolution of the simulation affects the turbulence scales stored in the database above the buffer layer.

\begin{figure}
    \centering
    \includegraphics{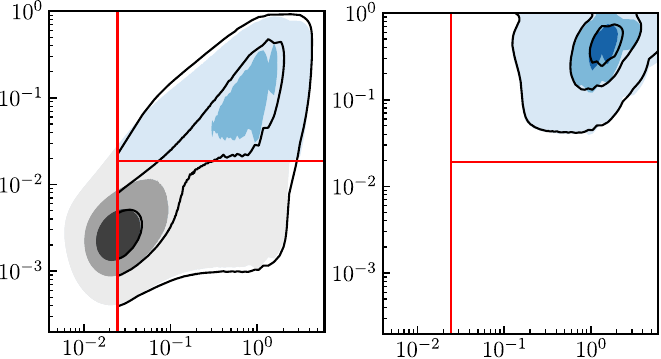}
    \vspace{0.4cm}%
    \mylab{-8.5cm}{-0.55cm}{$\lambda_3/h$}%
    \mylab{-3.0cm}{-0.55cm}{$\lambda_3/h$}%
    \mylab{-10.3cm}{5.4cm}{(a)}%
    \mylab{-4.65cm}{5.4cm}{(b)}%
    \mylab{-12.3cm}{3.55cm}{$x_2/h$}%
    \caption{(a) Premultiplied spectrum $k_3E_{11}$ of the streamwise velocity, for L5200 (shaded contours), and F5300 (line contours). (b) As \emph{a} but for the twofold premultiplied spectrum $yk_3E_{11}$. The red lines are $\lambda_3^+ \approx 65$ and $x_2^+ \approx 150$. The contours contain $10, 50$ and $90\%$ of the spectral mass of L5200, and the same absolute levels are used for F5300. \label{fig:spec2}}
\end{figure}

Finally, figure \ref{fig:spec2} shows the spectrum of the streamwise velocity as a function of the distance from the wall and the spanwise wavelength, where the ensemble average in \eqref{eq:spe} is extended to the streamwise wavenumber $k_1$. Compared with the full resolution spectrum of L5200, the spectrum of the retained scales of F5300,
is limited at $\lambda_3^+\gtrsim65$.
In figure \ref{fig:spec2}(a) we present the premultiplied $k_3E_{11}$ spectrum to show that half of the near-wall peak of the spectrum is not captured in the database.
If we consider that the low resolution affects the scales below $x_2^+ \lesssim 150$, indicated by a horizontal red line, then the near-wall peak, and seemingly an important fraction of the energy of $u_1$ is outside the scope of the database. However, when we consider the energy spectrum of $u_1$ premultiplied by $y$ and $k_3$, as shown in \ref{fig:spec2}(b), conclusions are very different. Usually the spectrum is only premultiplied in $k_3$ to observe both the near-wall and the outer peaks, but this representation is misleading in terms of the total energy contained in each peak. While the near-wall peak is very intense, it spans a short fraction of the channel and thus, the energy it contains is limited.
The twofold premultiplied spectrum, in which integrating across visually equivalent areas reflects the correct amount of energy contained in the wavenumbers beneath them, clearly shows that the database contains all the information of the very large, and most energetic scales of $u_1$.  

\subsection{Temporal statistics}

\begin{figure}
\centering
\includegraphics[width=\textwidth]{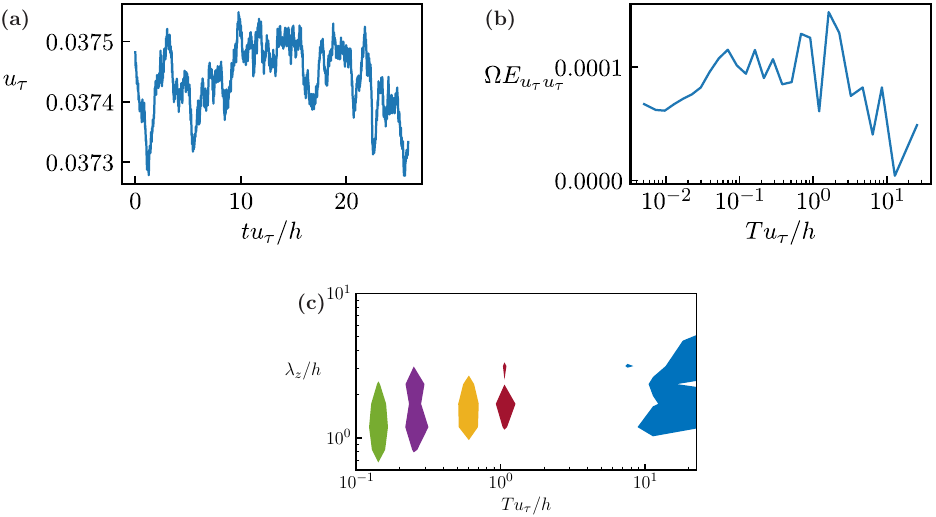}%
\mylab{-8cm}{0cm}{\colorbox{white}{$Tu_\tau/h$}}%
\mylab{-11.7cm}{2.45cm}{\colorbox{white}{$\lambda_3/h$}}%
\caption{(a) Time evolution of the friction velocity at the bottom wall. 
(b) Premultiplied temporal spectrum of the signal in (a). 
(c) Premultiplied $(x_3,t)$ spectra of the first streamwise Fourier modes of $u_1$ at $x_2/h = 0.4$. From right to left, $\lambda_1 = \infty$ (blue), $\lambda_1/h = 8\pi$ (red), $\lambda_1/h = 4\pi$ (yellow), $\lambda_1/h = 2\pi$ (purple), $\lambda_1 /h = \pi$ (green). The contour is 40\% of the maximum for each streamwise wavelength. F5000.
\label{fig:time}}
\end{figure}

Figure \ref{fig:time}(a) shows the temporal evolution of the instantaneous friction velocity
$u_\tau$ in the bottom wall, across approximately 25 \ett. We observe large variations of $u_\tau$ with a slow characteristic time, and small variations with a much faster time scale. 
The largest differences in the value of
$u_\tau$ within time windows of one \ett{} are approximately 5 times smaller than those within
10 \ett. To characterize the temporal structure of the friction velocity,  we calculate its premultiplied temporal spectrum, $\Omega E_{u_\tau u\tau}$, where $\Omega$ is the frequency, and show it against the period in figure \ref{fig:time}(b).
The spectrum is obtained by smoothly windowing the temporal signal to prevent its lack of periodicity from contaminating
the data \cite{welch1967use}.
An important fraction of the total friction velocity fluctuations are related to time-scales above one \ett{}, indicating that structures with longer lifetimes are present in the
flow, and that they contribute to the shear stress at the wall. Approximately, 34\% of the spectral mass, and thus of the skin friction fluctuations, is contained in periods longer than one \ett. 

Figure \ref{fig:time}(c)
shows the $(x_3,t)$ spectra of several streamwise Fourier modes at $x_2/h = 0.4$. At this height the streamwise velocity structures the are longest \cite{jim:2012}.
The two longest streamwise modes have characteristic times longer than one eddy turnover,
independently of their spanwise wavelength. Most of the energy of the infinitely long
streamwise streaks, represented by the $k_1=0$ wavenumber, is contained in lifetimes longer
than 6 \ett{}. These structures have a typical width of $\lambda_3/h \approx 2$-$3$, wider than
the typical size of the outer-layer streaks \cite{sil:jim:14}.
The structures
represented by the modes with $\lambda_1/h = 8\pi$ can be envisioned as streamwise velocity
streaks with a length of approximately $12h$ and a width of $1.5h$. Their temporal spectra \change{peak} at $Tu_\tau/h = 1$--$2$, indicating that they have shorter lifetimes than the infinitely long streamwise streaks. 
In any case, the temporal span of the database is a few times longer than the characteristic lifetimes of the largest structures in the channel,
proving that the database properly covers the dynamics of the large scales of the flow. 

%

\section{Conclusions\label{sec:conclusions}}

We have presented a new database of turbulent channel flow containing time-resolved
simulations at $Re_\tau = 2000$ and $Re_\tau = 5300$, for times up to $\ett \sim 30$, in
large boxes of size $(8\pi\times 3 \pi)h$. 
With an adequate spatio-temporal resolution to cover all the scales of the flow,
the size of the database becomes the critical factor, making the it impossible to store.
In the present case, it would require
approximately 60 PB. To alleviate these requirements, and since the database is mainly
intended to study the relatively large scales of the logarithmic layer and outer region, we only store
structures filtered above the dissipative range,
even if the simulation runs at full resolution.
To capture the influence of the discarded scales on the stored ones,
the database also includes the \change{time-resolved SGS stress tensor}, and the
filtered enstrophy. This is still more economical than storing the full flow fields ($\sim
150$ TB), and allows the reconstruction of other quantities of interest, such as the time
derivatives of the filtered velocities, the filtered pressure, and the filtered norm of the
rate-of-strain tensor. 
In this sense, the database is `second-order consistent', because any
observable that depends on quadratic velocity products can be computed \emph{a posteriori}. In
particular, this property, together with a relatively high Reynolds numbers, makes the
database potentially useful for \emph{a priori} testing and development of new LES models.

In order to achieve long running times, we slightly degrade the resolution of the computational grid without affecting the statistics of the retained scales. The first- and second-order, one- and two-point
statistics of the database corresponds to those of `healthy' logarithmic and outer regions. We show that this approach has potential applications
in other DNSs that target the large scales.
To further reduce the cost of the simulations, we employ a novel code which runs on many
distributed GPUs, and takes advantage of their highly efficient and powerful architectures.

From a preliminary analysis of the data, we show that time scales longer than one
eddy turnover time are present in the channel flow, especially for the streamwise velocity
component, and that dynamics associated with these time scales are responsible for about one third of the variation in the total turbulent skin friction.
We relate these fluctuations with with very long streamwise structures, whose lifetimes is at most $tu_\tau/h \approx 6$. The database has a long enough time history to cover a few of these lifetimes,
proving its suitability to study the large-scale dynamics of the channel flow. 
More detailed studies of the temporal statistics will be conducted in future works.

\section*{Acknowledgments}
This work was supported by the COTURB project of the European Research Council
(ERC2014.AdG-669505). The new simulations ran in PizDaint under the PRACE Tier-0 project
TREC.
We are grateful to M. Fatica and E. Phillips, from Nvidia, for invaluable help in optimising the performance of the MPI-CUDA code. We are also grateful to the anonymous referees of this manuscript who helped improving its clarity.

\def\appendixname{Appendix~} 
\appendix

\section{Algorithm and numerical set-up}
\label{sec:alg}

The formulation used to simulate the turbulent flow between two parallel planes with periodic boundary conditions 
is described in detail in \cite{kim1987turbulence}, and is to date a standard procedure. 
The Navier--Stokes equations are reduced to two equations for the mean profiles, $U_1$ and $U_3$, an equation for the Laplacian of the wall-normal velocity, $\chi=\partial_{kk} u_2$, and an equation for the wall-normal vorticity, ${\omega_2} = \partial_3 u_1 - \partial_1 u_3$. These last two equations read
\begin{equation}
\partial_t \chi=\mathcal{N}_{\chi}+\nu\partial_{kk}\chi,
\label{equ:v}
\end{equation}
\begin{equation}
\partial_t {\omega_2} =\mathcal{N}_{{\omega_2}}+\nu\partial_{kk} {\omega_2},
\label{equ:nabla}
\end{equation}
where $\nu$ is the molecular viscosity, and  
\begin{align}
\mathcal{N}_{\chi} &=-\partial_2(\partial_1 \mathcal{H}_1+\partial_3 \mathcal{H}_3)+(\partial_{11}+\partial_{33}) \mathcal{H}_2,\\
\mathcal{N}_{{\omega_2}} &=\partial_3 \mathcal{H}_1-\partial_1 \mathcal{H}_3,
\label{equ:hghv}
\end{align}
%
%
are the contributions of the non-linear terms to $\chi$ and ${\omega_2}$. Here $\boldsymbol{\mathcal{H}}=\boldsymbol{u}\times\boldsymbol{\omega}$, and $\boldsymbol \omega$ is the vorticity vector.
These equations present the advantage of hiding the problematic pressure term, but require the solution of
a fourth-order equation on $u_2$, with prescribed boundary conditions $u_2=0$ and $\partial_2 u_2=0$ at $x_2=(0, 2h)$.

In both periodic directions, (\ref{equ:v}) and (\ref{equ:nabla}) are projected on a Fourier basis with $N_1/2$ and $N_3/2$ modes and wavenumbers $k_1$ and $k_3$,  
which results in a uniformly spaced mesh of $N_1$ and $N_3$ collocation nodes in real space. The number of points, $N_1$ and $N_3$, are chosen to keep the desired resolution.
The non-linear terms are computed with a fully dealiased pseudo-spectral method \cite[see][]{can:hus:qua:zan:88}.
In the wall-normal direction equations are discretized on a non-uniform mesh of $N_2$ points, specifically adjusted to keep an adequate resolution at every wall distance. 
Derivatives in that direction are computed directly on the mesh using compact finite differences with spectral-like resolution \cite{lel:92,gamet1999compact}.
The compact finite differences are approximated with stencils of seven points in the centre of the domain, and adapted close to the wall. In all cases, they are designed to achieve maximum order.
A semi-implicit third order Runge--Kutta (RK3) is used for temporal integration \cite{spalart1991spectral}. Time-integration of the viscous terms is split in an explicit and an implicit part. The $n$-th step of the RK3 integrator for $\chi$  reads
\begin{align}
(1-\Delta t\beta\nu\partial_{kk})\chi^{n+1}&=\chi^{n}+\Delta t(\alpha\nu\partial_{kk}\chi^{n}+\gamma\mathcal{N}^{n}_\chi+\zeta\mathcal{N}^{n-1}_\chi), \label{chi1} \\
\partial_{kk} u^{n+1}&=\chi^{n+1}, \label{chi2}
\end{align}
with boundary conditions $u_2^{n+1}=0$, $\partial_2 u_2^{n+1}=0$ at $x_2= (0, 2h)$. Similarly, for ${\omega_2}$ we have
\begin{align}
(1-\Delta t\beta\nu\partial_{kk}){\omega_2}^{n+1}&={\omega_2}^{n}+\Delta t(\alpha\nu\partial_{kk}{\omega_2}^{n}+\gamma\mathcal{N}^{n}_{\omega_2}+\zeta\mathcal{N}^{n-1}_{\omega_2}),\label{psi}
\end{align}
with boundary conditions ${\omega_2}^{n+1}=0$ at $x_2= (0, 2h)$. Here $\Delta t$ is the time step, and $\alpha,\beta,\gamma$ and $\zeta$ are the RK3 coefficients \cite[][]{spalart1991spectral}.
Eqs. (\ref{chi1}) and (\ref{chi2}) conform a fourth order differential equation on $u_2$, which is solved by obtaining particular and homogeneous solutions of (\ref{chi1}),
and then combining them to satisfy the boundary conditions of  $\partial_2 u_2$ \citep[][]{kim1987turbulence}. 

\section{Enforcing boundary conditions at high Reynolds numbers}
\label{sec:enf}

High-order, high-resolution compact finite difference approximations for the derivatives in the wall-normal direction are frequently used in DNSs of turbulent channel flows,
mainly because they offer very 
good resolution properties at an affordable cost.
However, these methods are
prone to numerical errors if they are not developed together with the mesh. 
The most numerically sensitive operation in the vorticity--Laplacian formulation of the
Navier--Stokes equations is the solution of the biharmonic operator in (\ref{chi1}--\ref{chi2}),
which requires the solution of multiple homogeneous Helmholtz equations to impose the boundary
conditions. 
For high Reynolds numbers and small wavelengths, these solutions include very thin boundary layers near the wall 
which are difficult to reproduce numerically. 
These boundary layers need not be faithfully resolved at all wavenumbers, since they only affect the very viscous layer close to the wall, 
but, numerical errors in these solutions can compromise the full stability of the numerical simulations.

\begin{figure}
    \centering
    \includegraphics[trim=0.42cm 0cm 0cm 0cm, clip, width=7.9cm]{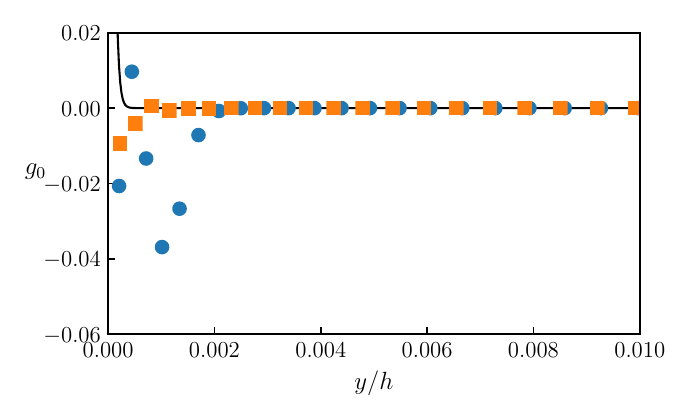}\hspace{0.4cm}%
    \raisebox{2.8cm}{%
    \begin{minipage}{8cm}%
    \hfill Matrix index\hfill\phantom{.}\\
    \includegraphics[width=0.49\textwidth]{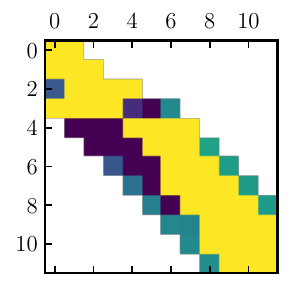}\hspace{-6pt}%
    \includegraphics[width=0.49\textwidth]{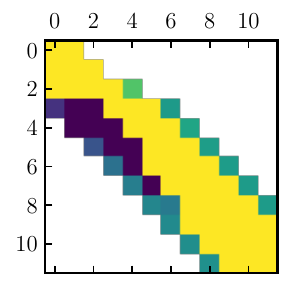}\\[-6.5pt]%
    \includegraphics[width=0.49\textwidth]{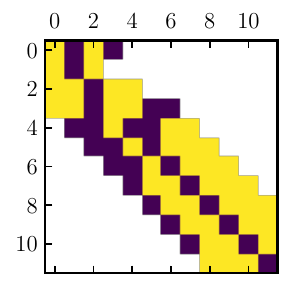}\hspace{-6pt}%
    \includegraphics[width=0.49\textwidth]{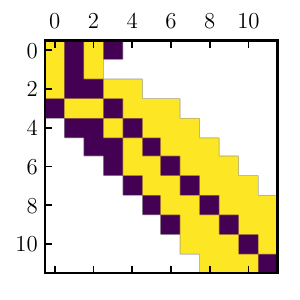}%
    \end{minipage}%
    }\mylab{-8.3cm}{1.5cm}{\rotatebox{90}{Matrix index}}%
    \mylab{-12.4cm}{5cm}{\small (a)}\mylab{-12.5cm}{0.3cm}{\scalebox{0.75}{\colorbox{white}{$x_2/h$}}}%
    \mylab{-5cm}{5.55cm}{\small (b)}%
    \mylab{-1.4cm}{5.55cm}{\small (c)}%
    \mylab{-5cm}{1.8cm}{\small (d)}%
    \mylab{-1.4cm}{1.8cm}{\small (e)}%
    \caption{(a) Solution of the Helmholtz equation for the most critical wavenumber. -------- Analytic solution. \textcolor{mlib0}{\fullcirc}, original mesh; \textcolor{mlib1}{\fullsquare},  optimised mesh. (b) First 10-by-10 elements of $A_\text{std}$. (c) First 10-by-10 elements of $A_\text{opt}$. (d) First 10-by-10 elements of $B_\text{std}$. (e) First 10-by-10 elements of $B_\text{opt}$. (b-d) The yellow coefficients are positive, and the blue coefficients are negatives.}
    \label{fig:mesh}
\end{figure}

We report critical numerical errors in the
homogeneous solutions of the biharmonic operator when discretized using compact finite differences of up to twelfth order of consistency.
Figure \ref{fig:mesh} compares the worst case solutions of the biharmonic operator used in the simulation (\ref{chi1}--\ref{chi2}). 
The grid spacing of the original mesh,
represented in the figure by circles, 
are optimised considering only physical arguments, 
as in \cite{hoy:jim:2006}.

These solutions are obtained as follows.
First, the approximation to the second derivative is obtained as the solution to the linear problem,
\begin{equation}
    \mat{A}\partial_{22}\zeta = \mat{B}\zeta,
\end{equation}
where $\zeta$ is a generic function of $x_2$ evaluated in the mesh points, and $\mat{A}$ and $\mat{B}$ are $N_2$-by-$N_2$ matrices. 
Computational efficiency requires that $\mat{A}$ and $\mat{B}$ are $N$-banded matrices, where $N=7$ in our case.
For derivatives computed directly on a non-uniform grid \citep{gamet1999compact}, $\mat A$ and $\mat B$ are obtained by solving $N_2$ 
linear systems of fourteen equations, one system for each grid point $(x_2)_i$,
\begin{gather}
    \sum_{j=-3}^{j=3}B_{i(i+j)}\deltah_{ij}^k - \mH(k-2)\frac{k!}{(k-2)!} \sum_{j=-3}^{j=3}A_{i(i+j)}\deltah_{ij}^{k-2} = \mH(k-2)\mH(2-k),\quad k=0,2\dots,13;\label{eq:diff}\\
    \deltah_{ij} = (x_2)_{i+j} - (x_2)_i,\\
    \mH(n) =
    \begin{cases}
        0 &  n <0,\\
        1 & \text{otherwise.}
    \end{cases}\label{eq:difflast}
\end{gather}
In order to reduce the condition number of the system, \eqref{eq:diff} are premultiplied by $\deltah_{i2}^{2-k}$ before obtaining its solution.
Let us note that \eqref{eq:diff} is not valid for the corner points, where the stencils are reduced progressively.
The compact finite differences approximation, 
together with a projection of the flow field on a Fourier basis in the wall-normal directions, transforms the homogeneous part of (\ref{chi1}) into a discrete Helmholtz equation,
\begin{gather}
    (\mat{B}-  k_\star\mat{A})g = 0,\label{eq:modpoisson}
\end{gather}
where $k_\star = k_1^2+k_3^2 + \beta/(\nu\Delta t)$ is a modified wavenumber which includes the effect of the RK3 integrator, and $g$ is a numerical approximation to the homogeneous solution of (\ref{chi1}). 
This equation requires Dirichlet boundary conditions $(g)_0 = 1$, and $(g)_{N_2} = 0$.
A mesh obtained with physical criteria \cite{hoy:jim:2006} worked well 
for $N_2=512$ at $Re_{tau}=2000$, but, as shown in figure \ref{fig:mesh},
led to large errors in the first few grid points for $N_2=1024$ and $Re_\tau=5300$. In some cases, these errors were enough to make the simulation unstable.
We obtained an optimal grid that eliminates the numerical instability by slightly displacing the first 20 grid points, without affecting too much the individual grid spacing, thus preserving the physical properties of the mesh. 
The optimal grid is obtained by minimizing a prescribed functional,
\begin{equation}
    L_g(\mu_\text{opt}; \mathcal{C}) = \left\| g_\text{opt} - g_\text{ana}\right\| + \mathcal{C}\left\| \mu_\text{opt} - \mu_\text{ref}\right\|,
\end{equation}
with a steepest descent method. Here $g_\text{ana}$ is the analytic solution of \eqref{eq:modpoisson}, obtained as a combination of hyperbolic functions, $\mu$ contains the position of the 21 grid points in $x_2$, 
and $\mathcal{C}$ is a penalisation coefficient that prevents large drifts of the grid respect to the physically meaningful values of the original grid.
The resulting optimal mesh produces twelve times less error in the homogeneous solutions than the original mesh.
The value of $\mathcal{C}$ was set to
$\mathcal{C} \approx 10^{-3}$,
but its exact value has minimal influence 
on the optimal mesh.
Figures \ref{fig:mesh}(b--d) offer some insight on the origin of the errors generated by the original grid.
The original mesh has a sudden change in the symmetry of the coefficients close to the domain edge, which is removed by the optimisation procedure.
We arbitrarily modify the grid to cause the position of the symmetry reversal to move up and down the mesh, and observe that the largest errors are always measured in its vicinity, indicating that this is the cause of the numerical \change{instability.}

\section{Domain decomposition and optimisation strategy}
\label{sec:opt}

The computational domain is decomposed in $x_2x_3$-planes across GPUs.
For the computation of the non-linear terms, $\boldsymbol{\mathcal{H}}$,  the three components of the velocity and two of the vorticity, $\omega_1$ and $\omega_3$, are transposed to $x_3x_1$-planes. This plane-to-plane decomposition is chosen to reduce communication operations among nodes. First, it saves one global transpose per evaluation of the non-linear term, because ${\omega_2}$ can be calculated directly from $u_1$ and $u_3$ in Fourier space when $x_{3}x_1$-planes are available.
Second, it reduces the size of the MPI messages in the GPU-to-GPU communications as the fields can be zero-padded for dealiasing after transposed.
A plane-pencil decomposition requires transposing $50\%$ larger buffers.  
On the other hand, this decomposition implies that GPUs must be able to accommodate at least full $x_{3}x_{1}$-planes and that the maximum number of GPUs is limited to $N_{2}$. This limitations are not critical if we consider the increasing memory and compute capabilities of GPUs, and that future high performance computing systems tend towards less but more powerful nodes with more available memory \cite{hines2018stepping}. 

In standard  DNS spectral codes, the most time-consuming part of the computations is the non-linear convolution, whose evaluation requires several Fourier transforms and global MPI transposes. The performance of the code is specially sensitive to the latter, which take between 30\% to 60\% of the total runtime, depending on the code and the network architecture \citep{borrell2013code, lee2013petascale}. 
Since MPI transposes are performed from the CPU memory, we also have to consider an additional overhead due to memory transfers between the GPU and the CPU. We focus most of the optimisation efforts on this part of the code, and exploit the capabilities of GPUs to achieve an efficient execution.
An important advantage of using GPUs is the possibility of asynchronously executing code while performing the MPI transposes. Here we present a well-planned strategy to achieve almost complete overlapping of computations and communications, yielding a significant speed-up.
This strategy can be easily adapted to other spectral DNS codes.


Concurrent GPU execution and memory transfer 
is achieved by setting three different streams on the device side: a computation stream, a device-to-host-copy stream (D2H) and a host-to-device-copy stream (H2D).
We control the correct execution of the different tasks using the synchronisation tools and event management functions available in the CUDA runtime API.

We optimise the evaluation of the non-linear terms, and build the rest of the RK3 sub-step around this optimisation.
In the implemented formulation, $u_2$ is available at the beginning of each sub-step, and $u_1$ and $u_3$ are calculated from  ${\omega_2}$ and $\partial_2 u_2$, which are also available.
Thus we start the computation of the RK3 sub-step by scheduling the global transpose of $u_2$, which is first pre-transposed on the GPU, and then copied to the host.
At the same time, both $u_1$ and $u_3$ are calculated from $\partial_2 u_2$ and ${\omega_2}$, and scheduled for global transpose. While $u_1$, $u_2$ and $u_3$ are being copied to the host and transposed to $x_1x_3$-planes using MPI all-to-all functions, the GPU is available to compute $\partial_2 u_1$ and $\partial_2 u_3$, which are required to calculate $\omega_1$ and $\omega_3$.
When  $\partial_2 u_1$ and $\partial_2 w_3$ are ready in the GPU, they are copied to the host and transposed into $x_{3}x_{1}$-planes. When $u_1$, $u_2$, $u_3$, $\partial_2 u_1$ and $\partial_2 u_3$ are copied back to the GPU, ${\omega_2}$, $\omega_1$ and $\omega_3$ are calculated, and the three components of the velocity and the vorticity are transformed to real space. All the compute, copy and transpose operations are performed concurrently.
Let us note that $\omega_2$ can be calculated in place after the transpose due to the plane-plane decomposition adopted. We are now ready to evaluate the non-linear terms in real space.
Only $\mathcal{H}_1$ and $\mathcal{H}_3$ are needed for $\mathcal{N}_{{\omega_2}}$, where $\mathcal{H}_1=u_2\omega_3-u_3{\omega_2}$ and $\mathcal{H}_3=u_1{\omega_2}-u_2\omega_1$. First we calculate $\mathcal{H}_1$ and $\mathcal{H}_3$, transform them to Fourier space and schedule their transpose back to $x_2x_3$-planes. 
Then we calculate $\mathcal{N}_{{\omega_2}}$.
A similar procedure is followed for $\mathcal{H}_3$ to calculate $\mathcal{N}_{\omega_2}$. 

These operations constitute the core of the code, and represent most of the total computation time in a full RK3 substep.
A few operations remain to complete a RK3 sub-step: computing second derivatives for the explicit part of the viscous terms (explicit linear operator), adding them together with the non-linear terms of the previous sub-step (adding $\mathcal N^{n-1}$), and taking the implicit steps of the viscosity (implicit linear operator). Except for the implicit steps, which must be completed at the end of the sub-step,
the rest of the operations can be distributed to fill the gaps when the GPU is idle waiting for copies and transposes to finish. The idle CPU time is also exploited to compute on the fly statistics, and the evolution of the mean velocity profiles.
In table \ref{tab:1} an schematic overview of a complete RK3 sub-step is conceptually represented. Dependencies, which are marked with colours, show how each block is arranged to obtain maximum concurrency.


 \begin{table}[]
\centering
\resizebox{0.9\textwidth}{!}{
\begin{tabular}{|l|l|l|l|}
\hline
\bf{Compute stream}                            & \bf{D2H stream}                        & \bf{H2D stream}                                   & \bf{Host stream}                \\ \hline
\cellcolor{blue!25}calculate $u_1$               & \cellcolor{green!25}copy $u_2$ to host   &                                                   &                            \\
\cellcolor{red!25}calculate $u_3$                & \cellcolor{blue!25}copy $u_1$ to host    &                                                   & \cellcolor{green!25}MPI transp. $u_2$              \\
\cellcolor{magenta!25}calculate $\partial_2 u_1$ & \cellcolor{red!25}copy $u_3$ to host     & \cellcolor{green!25}copy $u_2$ to device            & \cellcolor{blue!25}MPI transp. $u_1$              \\
\cellcolor{orange!25}calculate $\partial_2 u_3$& \cellcolor{magenta!25} copy $\partial_2 u_1$ to host           & \cellcolor{blue!25}copy $u_1$ to device             & \cellcolor{red!25}MPI transp. $u_3$     \\
calculate $\partial_{22} \chi$                 & \cellcolor{orange!25}copy $\partial_2 u_3$ to host & \cellcolor{red!25}copy $u_3$ to device            & \cellcolor{magenta!25}MPI transp. $\partial_2 u_1$             \\
calculate $\partial_{22} {\omega_2} $                &                             & \cellcolor{magenta!25}copy $\partial_2 u_1$ to device            &\cellcolor{orange!25} MPI transp. $\partial_2 u_3$ \\
\cellcolor{green!25}FT to real $u_2$                                 &                             & \cellcolor{orange!25}copy $\partial_2 u_3$ to device&  \\
\cellcolor{blue!25}FT to real $u_1$                                 &                             &  &                            \\
\cellcolor{red!25}FT to real $u_3$                                 &                             &                               &                            \\
calculate ${\omega_2}$ and FT to real       &                             &                               &                            \\
\cellcolor{magenta!25}calculate $\omega_1$ and FT to real       &                             &                               &                            \\
\cellcolor{orange!25}calculate $\omega_3$ and FT to real       &                             &                               & calculate statistics       \\
\cellcolor{lime!25}calculate $\mathcal{H}_1$ and FT to complex         &                             &                               &                            \\
\cellcolor{violet!25}calculate $\mathcal{H}_3$ and FT to complex         &\cellcolor{lime!25} copy $\mathcal{H}_1$ to host          &                               &                            \\
\cellcolor{brown!25}calculate $\mathcal{H}_2$ and FT to complex         &\cellcolor{violet!25} copy $\mathcal{H}_3$ to host          &                               &\cellcolor{lime!25} MPI transp. $\mathcal{H}_1$           \\
update RHS of $\chi$                  &\cellcolor{brown!25} copy $H_2$ to host          &\cellcolor{lime!25} copy $\mathcal{H}_1$ to device          &\cellcolor{violet!25} MPI transp. $\mathcal{H}_3$           \\
update RHS of ${\omega_2}$                      &                             & \cellcolor{violet!25}copy $\mathcal{H}_3$ to device          &\cellcolor{brown!25} MPI transp. $\mathcal{H}_2$           \\
calculate $\mathcal{N}_{{\omega_2}}$             &                             &\cellcolor{brown!25} copy $\mathcal{H}_2$ to device          &                            \\
implicit step for ${\omega_2}$              &                             &                               &                            \\
calculate $\mathcal{N}_{\chi}$           &                             &                               &                            \\
implicit step for $\chi$            &                             &                               &                            \\ \hline
\end{tabular} 
}
\caption{ Conceptual representation of the arrangements of operations in the compute, copy, D2H, H2D and host streams for the computation of a sub-step of the RK3 temporal integrator. FT denotes Fourier transform.
Color indicates dependencies.}
 \label{tab:3}
\end{table}

\begin{figure}
\centering
    \includegraphics[width=\textwidth]{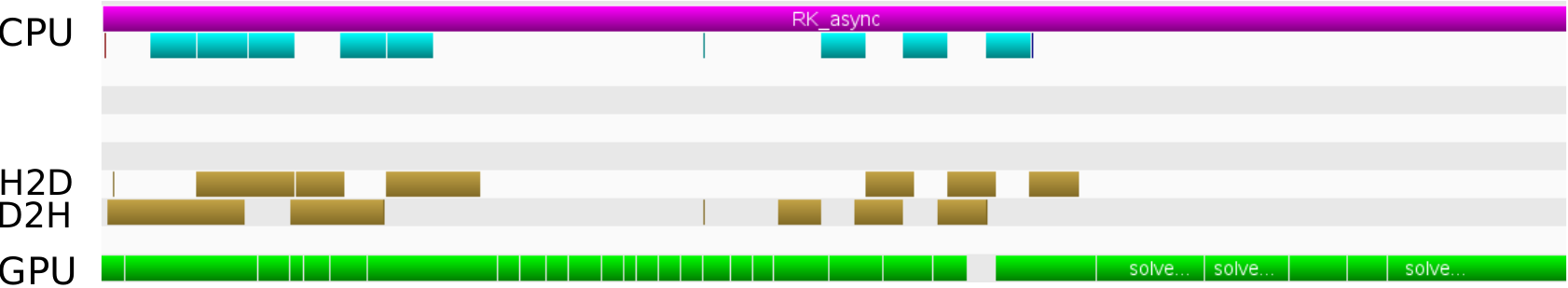}
\caption{Execution profile of a RK3 sub-step of a $1024\times256\times1024$ case \change{run} on 32 GPUs in Minotauro at BSC. In blue MPI transposes, in brown D2H and H2D copies, and in green GPU execution. In this case, we have obtained a complete overlapping of GPU execution and communications.\label{fig:prof}} 

\end{figure}

In figure \ref{fig:prof} we show a real compute profile of the RK3 sub-step obtained using the
CUDA visual profiler in Minotauro\change{, a GPU HPC Cluster at the Barcelona Supercomputing Centre}. A total overlapping of GPU computations, device-host memory transfer and MPI/CPU execution is achieved.
From the point of view of the computations,
the code performs as if running on a large GPU with a unique shared memory.
In this case, the time spent in MPI transposes is $24\%$ of the total run-time, while GPU/CPU memory transfer accounts for $49\%$. This is a specially convenient situation, in which the network bandwidth and the GPU-CPU bandwidth is comparable.
This optimisation strategy is able to achieve a considerable speed-up with respect to a synchronous code, even if the GPU/CPU transfer is not considered, but its performance depends strongly on the balance between the bandwidth of the network, and the bandwidth and compute capabilities of the GPUs. 



\section{Scaling of the code}
\label{sec:scal}

The scaling of the code has been tested for a variety of cases.
In all cases the grid has an 
aspect ratio similar to the largest production run. 
The maximum and minimum number of GPUs used to run each case is limited by code design and by GPU memory.
The plane-plane domain decomposition is intended to exploit maximum GPU occupancy and to minimize global
communications, but constrains the maximum number
of GPUs on which the code can run to $N_2$.
When both numbers
coincide, a single $x_3x_1$-plane per buffer is stored in each GPU. 
The minimum number of GPUs is constrained by the available device memory.
A summary of the test cases is presented in table \ref{tab:1}.

\begin{table}
\begin{center}
\begin{tabular}{c c c c } 
 \hline
 &$N_1\times N_2\times N_3$ & $N^\text{min}_\text{gpus}-N^\text{max}_\text{gpus}$ & $\tau_\text{min}$ (\text{ns})  \\ 
 \hline
 {\color{red}$\star$}      &  $1024\times256\times1024\ $ & $16-256$      & $60$  \\ 
 {\color{blue}$+$}         &  $2048\times512\times2048\ $ & $64-512$      & $63$  \\
 {\color{magenta}$\circ$}  &     $4096\times1024\times4096$ & $512-1024$  & $65$  \\
 $\square$& $6144\times1024\times4096$ & $512-1024$          & $63$  \\
 \hline
\end{tabular}

\caption{Parameters of the test cases run for the scalability analysis, where $\tau$ is the
time per step and degree of freedom (see text for definition).
$N^\text{min}_\text{gpus}$ and $N^\text{max}_\text{gpus}$ are the minimum and maximum number of GPUs in which
each case has been run. $N_\text{gpus}$ is always chosen a power of two. \label{tab:1}}
\end{center}
\end{table}

\begin{figure}[ht]
\centering
%
\includegraphics[width=\textwidth]{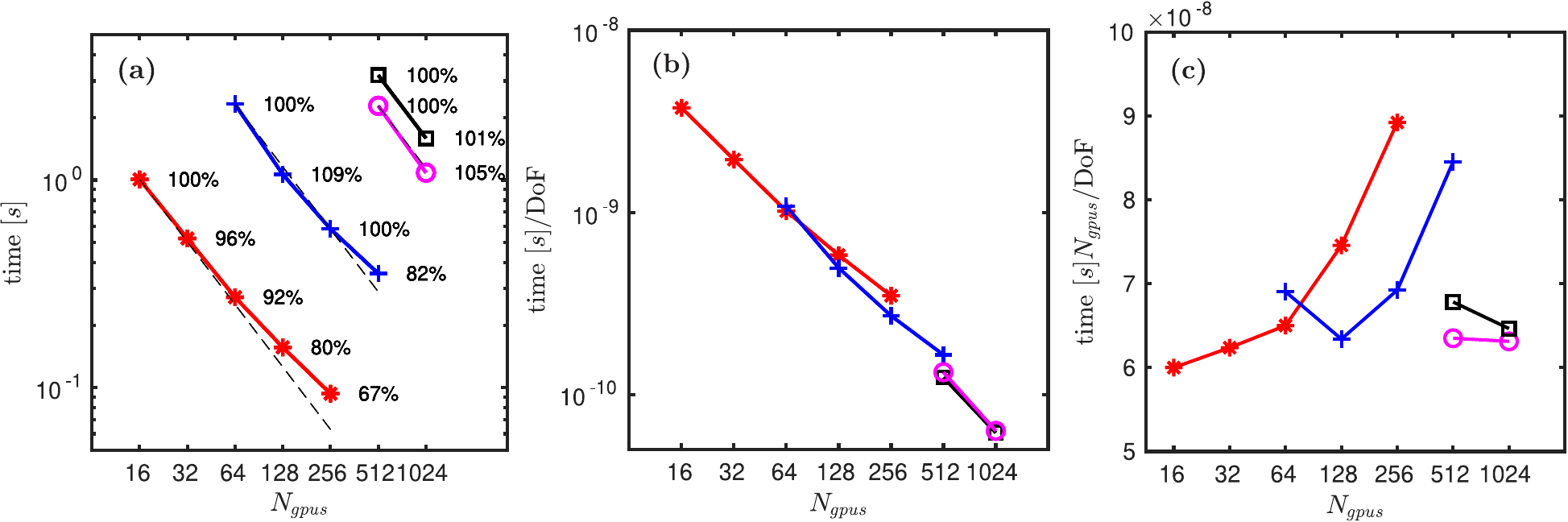}%

\caption{(a) Execution time of one RK3 step (seconds) as a function of the number of GPUs for
different test cases. Percentages represent efficiency, $t/t_\text{ideal}$.
(b) Execution time of one RK3 step per degree of freedom. 
(c)Time of a RK3 step per degree of freedom compensated with the number of GPUs.\label{fig:21}} 
\end{figure}

Scaling tests have been conducted in PizDaint at CSCS, a GPU-based supercomputer part of the
PRACE Tier-0 network \cite{website:cscs}. 
Results of the strong scaling are shown in figure \ref{fig:21}(a).
The efficiency is defined as $t/t_\text{ideal}$, where $t$ is the measured execution time per RK3 step in $N_\text{gpus}$,
and $t_\text{ideal}=t_0N_0/N_\text{gpus}$ is the ideal execution time.
Here $N_0$ is the minimum number of GPUs that can be used for each grid size, and $t_0$ is
the measured execution time for that case. The efficiency of the medium and large cases is very close to {100\%}, with super-linear scaling in some cases.
In figures \ref{fig:21}(b, c), we show results of the weak scaling. The RK3 step run-time per GPU and degree of freedom is very similar in all cases.

\newcommand{\doi}[1]{\url{http://doi.org/#1}}
\bibliographystyle{unsrtnat}
\bibliography{report}

\begin{thebibliography}{48}
\providecommand{\natexlab}[1]{#1}
\providecommand{\url}[1]{\texttt{#1}}
\expandafter\ifx\csname urlstyle\endcsname\relax
  \providecommand{\doi}[1]{doi: #1}\else
  \providecommand{\doi}{doi: \begingroup \urlstyle{rm}\Url}\fi

\bibitem[Jim{\'e}nez(2013)]{jimenez2013near}
J.~Jim{\'e}nez.
\newblock Near-wall turbulence.
\newblock \emph{Phys. Fluids}, 25:\penalty0 101302, 2013.

\bibitem[Kim et~al.(1987)Kim, Moin, and Moser]{kim1987turbulence}
J.~Kim, P.~Moin, and R.~Moser.
\newblock Turbulence statistics in fully developed channel flow at low
  {R}eynolds number.
\newblock \emph{J. Fluid Mech.}, 177:\penalty0 133--166, 1987.

\bibitem[Moser et~al.(1999)Moser, Kim, and Mansour]{mos:kim:man:99}
R.~D. Moser, J.~Kim, and N.~N. Mansour.
\newblock Direct numerical simulation of turbulent channel flow up to
  ${Re}_\tau= 590$.
\newblock \emph{Phys. Fluids}, 11\penalty0 (4):\penalty0 943--945, 1999.
\newblock \doi{10.1063/1.869966}.
\newblock URL \url{http://link.aip.org/link/?PHF/11/943/1}.

\bibitem[del \'Alamo and Jim\'enez(2003)]{ala:jim:03}
J.~C. del \'Alamo and J.~Jim\'enez.
\newblock Spectra of the very large anisotropic scales in turbulent channels.
\newblock \emph{Phys. Fluids A}, 15:\penalty0 L41--L44, 2003.

\bibitem[Hoyas and Jim{\'e}nez(2006)]{hoy:jim:2006}
S.~Hoyas and J.~Jim{\'e}nez.
\newblock Scaling of the velocity fluctuations in turbulent channels up to
  ${Re}_\tau= 2003$.
\newblock \emph{Phys. Fluids}, 18\penalty0 (1):\penalty0 011702, 2006.
\newblock \doi{10.1063/1.2162185}.
\newblock URL \url{http://link.aip.org/link/?PHF/18/011702/1}.

\bibitem[Lee and Moser(2015)]{leemoser15}
M.~Lee and R.~D. Moser.
\newblock Direct numerical simulation of turbulent channel flow up to
  ${R}e_\tau\approx 5200$.
\newblock \emph{J. Fluid Mech.}, pages 395--415, 2015.

\bibitem[Jim{\'e}nez and Moin(1991)]{jim:moi:91}
J.~Jim{\'e}nez and P.~Moin.
\newblock The minimal flow unit in near-wall turbulence.
\newblock \emph{J. Fluid Mech.}, 225:\penalty0 213--240, 4 1991.
\newblock ISSN 1469-7645.
\newblock \doi{10.1017/S0022112091002033}.
\newblock URL \url{http://journals.cambridge.org/article_S0022112091002033}.

\bibitem[Hamilton et~al.(1995)Hamilton, Kim, and Waleffe]{ham:kim:wal:95}
J.~M. Hamilton, J.~Kim, and F.~Waleffe.
\newblock Regeneration mechanisms of near-wall turbulence structures.
\newblock \emph{J. Fluid Mech.}, 287:\penalty0 317--348, 1995.

\bibitem[Pope(2000)]{pop:00}
S.~B. Pope.
\newblock \emph{Turbulent flows}.
\newblock Cambridge U. Press, 2000.

\bibitem[Jim{\'e}nez(2012)]{jim:2012}
J.~Jim{\'e}nez.
\newblock Cascades in wall-bounded turbulence.
\newblock \emph{Ann. Rev. Fluid Mech.}, 44:\penalty0 27--45, 2012.
\newblock \doi{10.1146/annurev-fluid-120710-101039}.

\bibitem[Bernardini et~al.(2014)Bernardini, Pirozzoli, and
  Orlandi]{ber:pir:orl:14}
M.~Bernardini, S.~Pirozzoli, and P.~Orlandi.
\newblock Velocity statistics in turbulent channel flow up to ${R}e_\tau=4000$.
\newblock \emph{J. Fluid Mech.}, 742:\penalty0 171--191, 3 2014.
\newblock ISSN 1469-7645.
\newblock \doi{10.1017/jfm.2013.674}.
\newblock URL \url{http://journals.cambridge.org/article_S0022112013006745}.

\bibitem[Lozano-Dur\'an and Jim\'enez(2014)]{loz:jim:2014}
A.~Lozano-Dur\'an and J.~Jim\'enez.
\newblock Effect of the computational domain on direct simulations of turbulent
  channels up to ${Re}_\tau=4200$.
\newblock \emph{Phys. Fluids}, 26\penalty0 (1):\penalty0 011702, 2014.
\newblock \doi{10.1063/1.4862918}.

\bibitem[Yamamoto and Tsuji(2018)]{yamamoto2018numerical}
Y.~Yamamoto and Y.~Tsuji.
\newblock Numerical evidence of logarithmic regions in channel flow at
  ${R}e_\tau= 8000$.
\newblock \emph{Phys. Rev. Fluids}, 3:\penalty0 012602, 2018.

\bibitem[Jim{\'e}nez(1998)]{jim:98}
J.~Jim{\'e}nez.
\newblock The largest scales of turbulence.
\newblock In \emph{CTR Ann. Res. Briefs}, pages 137--154. Stanford Univ., 1998.

\bibitem[Kim and Adrian(1999)]{kim:adr:99}
K.~C. Kim and R.~J. Adrian.
\newblock Very large-scale motion in the outer layer.
\newblock \emph{Phys. Fluids}, 11\penalty0 (2):\penalty0 417--422, 1999.

\bibitem[del \'Alamo and Jim\'enez(2001)]{ala:jim:01}
J.~C. del \'Alamo and J.~Jim\'enez.
\newblock Direct numerical simulation of the very large anisotropic scales in a
  turbulent channel.
\newblock In \emph{{CTR} {A}nn. {R}es. {B}riefs}, pages 329--341. Stanford
  University, 2001.

\bibitem[Liu et~al.(2001)Liu, Adrian, and Hanratty]{liu:adr:han:01}
Z.~Liu, R.~J. Adrian, and T.~J. Hanratty.
\newblock Large-scale modes of turbulent channel flow: transport and structure.
\newblock \emph{J. Fluid Mech.}, 448:\penalty0 53--80, 2001.

\bibitem[del \'Alamo et~al.(2004)del \'Alamo, Jim\'enez, Zandonade, and
  Moser]{ala:jim:zan:mos:04}
J.~C. del \'Alamo, J.~Jim\'enez, P.~Zandonade, and R.~D. Moser.
\newblock Scaling of the energy spectra of turbulent channels.
\newblock \emph{J. Fluid Mech.}, 500:\penalty0 135--144, 2004.

\bibitem[Flores and Jim{\'e}nez(2010)]{flo:jim:10}
O.~Flores and J.~Jim{\'e}nez.
\newblock Hierarchy of minimal flow units in the logarithmic layer.
\newblock \emph{Phys. Fluids}, 22:\penalty0 071704, 2010.

\bibitem[Robinson(1991)]{rob:91}
S.~K. Robinson.
\newblock Coherent motions in the turbulent boundary layer.
\newblock \emph{Ann. Rev. Fluid Mech.}, 23\penalty0 (1):\penalty0 601--639,
  1991.
\newblock \doi{10.1146/annurev.fl.23.010191.003125}.
\newblock URL
  \url{http://www.annualreviews.org/doi/abs/10.1146/annurev.fl.23.010191.003125}.

\bibitem[{C}omputational {F}luid~{D}ynamics {L}ab.(2009)]{website:dmt}
UPM {C}omputational {F}luid~{D}ynamics {L}ab.
\newblock {C}omputational {F}luid {D}ynamics {L}ab.
\newblock \url{http://torroja.dmt.upm.es/}, May 2009.

\bibitem[Jim{\'{e}}nez(2018)]{summerprogram}
J.~Jim{\'{e}}nez.
\newblock Third {M}adrid summer school on turbulence.
\newblock \emph{J. Phys. Conf. Ser.}, 1001:\penalty0 011001, apr 2018.
\newblock \doi{10.1088/1742-6596/1001/1/011001}.
\newblock URL \url{https://doi.org/10.1088%2F1742-6596%2F1001%2F1%2F011001}.

\bibitem[Perlman et~al.(2007)Perlman, Burns, Li, and Meneveau]{pha_jhu}
E.~Perlman, R.~Burns, Y.~Li, and C.~Meneveau.
\newblock Data exploration of turbulence simulations using a database cluster.
\newblock In \emph{Proc. SC07}, pages 23.1--23.11. ACM, New York, 2007.
\newblock \doi{10.1145/1362622.1362654}.
\newblock {http://turbulence.pha.jhu.edu}.

\bibitem[Lozano-{D}ur\'{a}n and Jim\'enez(2014)]{lozano-time}
A.~Lozano-{D}ur\'{a}n and J.~Jim\'enez.
\newblock Time-resolved evolution of coherent structures in turbulent channels:
  characterization of eddies and cascades.
\newblock \emph{J. Fluid Mech.}, 759:\penalty0 432--471, 2014.

\bibitem[Hwang and Cossu(2011)]{hwa:cos:11}
Y.~Hwang and C.~Cossu.
\newblock Self-sustained processes in the logarithmic layer of turbulent
  channel flows.
\newblock \emph{Phys. Fluids}, 23:\penalty0 061702, 2011.

\bibitem[Mizuno and Jim\'enez(2013)]{miz:jim:2013}
Y.~Mizuno and J.~Jim\'enez.
\newblock Wall turbulence without walls.
\newblock \emph{J. Fluid Mech.}, 723:\penalty0 429--455, 5 2013.
\newblock ISSN 1469-7645.
\newblock \doi{10.1017/jfm.2013.137}.

\bibitem[Dong et~al.(2017)Dong, Lozano-Dur{\'a}n, Sekimoto, and
  Jim{\'e}nez]{dong17}
S.~Dong, A.~Lozano-Dur{\'a}n, A.~Sekimoto, and J.~Jim{\'e}nez.
\newblock Coherent structures in statistically stationary homogeneous shear
  turbulence.
\newblock \emph{J. Fluid Mech.}, 816:\penalty0 167--208, 2017.

\bibitem[Scovazzi et~al.(2001)Scovazzi, Jim\'enez, and Moin]{scov2001}
G.~Scovazzi, J.~Jim\'enez, and P.~Moin.
\newblock {LES} of the very large scales in a ${Re}_\tau=920$ channel.
\newblock In \emph{Proc. Div. Fluid Dyn.}, pages KF--5. Am. Phys. Soc., 2001.

\bibitem[Ishihara et~al.(2009)Ishihara, Gotoh, and Kaneda]{ishihara2009study}
T.~Ishihara, T.~Gotoh, and Y.~Kaneda.
\newblock Study of high--{R}eynolds number isotropic turbulence by direct
  numerical simulation.
\newblock \emph{Annu. Rev. Fluid Mech.}, 41:\penalty0 165--180, 2009.

\bibitem[Khajeh-Saeed and Perot(2013)]{kha:per:13}
A.~Khajeh-Saeed and J.~B. Perot.
\newblock Direct numerical simulation of turbulence using {GPU} accelerated
  supercomputers.
\newblock \emph{J. Comput. Phys.}, 235\penalty0 (0):\penalty0 241--257, 2013.
\newblock ISSN 0021-9991.
\newblock \doi{10.1016/j.jcp.2012.10.050}.
\newblock URL
  \url{http://www.sciencedirect.com/science/article/pii/S0021999112006547}.

\bibitem[Karantasis et~al.(2014)Karantasis, Polychronopoulos, and
  Ekaterinaris]{kar:pol:eka:14}
K.~I. Karantasis, E.~D. Polychronopoulos, and J.~A. Ekaterinaris.
\newblock High order accurate simulation of compressible flows on {GPU}
  clusters over {S}oftware {D}istributed {S}hared {M}emory.
\newblock \emph{Comput. Fluids}, 93\penalty0 (0):\penalty0 18--29, 2014.
\newblock ISSN 0045-7930.
\newblock \doi{10.1016/j.compfluid.2014.01.005}.
\newblock URL
  \url{http://www.sciencedirect.com/science/article/pii/S0045793014000127}.

\bibitem[Zhu et~al.(2018)Zhu, Phillips, Spandan, Donners, Ruetsch, Romero,
  Ostilla-M{\'o}nico, Yang, Lohse, Verzicco, F., and M.]{zhu2018afid}
X.~Zhu, E.~Phillips, V.~Spandan, J.~Donners, G.~Ruetsch, J.~Romero,
  R.~Ostilla-M{\'o}nico, Y.~Yang, D.~Lohse, R.~Verzicco, Massimiliano F., and
  Stevens R. J.~A. M.
\newblock {AFiD-GPU}: a versatile {N}avier--{S}tokes solver for wall-bounded
  turbulent flows on gpu clusters.
\newblock \emph{Comput. Phys. Commun.}, 229:\penalty0 199--210, 2018.

\bibitem[Vela-Mart\'in(2019)]{website:code}
Alberto Vela-Mart\'in.
\newblock \url{https://github.com/albertovelam/CHANNEL_GPU}, 2019.

\bibitem[Canuto et~al.(1988)Canuto, Hussaini, Quarteroni, and
  Zang]{can:hus:qua:zan:88}
C.~Canuto, M.~Y. Hussaini, A.~Quarteroni, and T.~A. Zang.
\newblock \emph{Spectral Methods in Fluid Dynamics}.
\newblock Springer-Verlag, Heidelberg, 1988.

\bibitem[Flores and Jim{\'e}nez(2006)]{flo:jim:06}
O.~Flores and J.~Jim{\'e}nez.
\newblock Effect of wall-boundary disturbances on turbulent channel flows.
\newblock \emph{J. Fluid Mech.}, 566:\penalty0 357--376, 2006.

\bibitem[Lele(1992)]{lel:92}
S.~K. Lele.
\newblock Compact finite difference schemes with spectral-like resolution.
\newblock \emph{J. Comput. Phys.}, 103\penalty0 (1):\penalty0 16--42, 1992.
\newblock ISSN 0021-9991.
\newblock \doi{10.1016/0021-9991(92)90324-R}.
\newblock URL
  \url{http://www.sciencedirect.com/science/article/pii/002199919290324R}.

\bibitem[Gamet et~al.(1999)Gamet, Ducros, Nicoud, and
  Poinsot]{gamet1999compact}
L.~Gamet, F.~Ducros, F.~Nicoud, and T.~Poinsot.
\newblock Compact finite difference schemes on non-uniform meshes.
  {A}pplication to direct numerical simulations of compressible flows.
\newblock \emph{Int. J. Numer. Meth. Fl.}, 29\penalty0 (2):\penalty0 159--191,
  1999.

\bibitem[Spalart et~al.(1991)Spalart, Moser, and Rogers]{spalart1991spectral}
P.~R Spalart, Robert~D. Moser, and M.~M. Rogers.
\newblock Spectral methods for the {N}avier--{S}tokes equations with one
  infinite and two periodic directions.
\newblock \emph{J. Comp. Phys.}, 96\penalty0 (2):\penalty0 297--324, 1991.

\bibitem[NVIDIA(2019)]{website:cufft}
NVIDIA.
\newblock \url{https://docs.nvidia.com/cuda/cufft/index.html}, 2019.

\bibitem[Kolmogorov(1941)]{kol:41}
A.~N. Kolmogorov.
\newblock {The Local Structure of Turbulence in Incompressible Viscous Fluid
  for Very Large {R}eynolds' Numbers}.
\newblock In \emph{Dokl. Akad. Nauk SSSR}, volume~30, pages 301--305, 1941.

\bibitem[Kim(1989)]{kim1989structure}
J.~Kim.
\newblock On the structure of pressure fluctuations in simulated turbulent
  channel flow.
\newblock \emph{J. Fluid Mech.}, 205:\penalty0 421--451, 1989.

\bibitem[Jim{\'e}nez(2004)]{jim:rou:04}
J.~Jim{\'e}nez.
\newblock Turbulent flows over rough walls.
\newblock \emph{Annu. Rev. Fluid Mech.}, 36:\penalty0 173--196, 2004.

\bibitem[Welch(1967)]{welch1967use}
P.~Welch.
\newblock The use of fast {F}ourier transform for the estimation of power
  spectra: a method based on time averaging over short, modified periodograms.
\newblock \emph{IEEE Trans. Audio Electroacoust.}, 15:\penalty0 70--73, 1967.

\bibitem[Sillero et~al.(2014)Sillero, Jim\'enez, and Moser]{sil:jim:14}
J.~Sillero, J.~Jim\'enez, and R.~D. Moser.
\newblock Two-point statistics for turbulent boundary layers and channels at
  {R}eynolds numbers up to $\delta^+ \sim 2000$.
\newblock \emph{Phys. Fluids}, 26:\penalty0 105--109, 2014.

\bibitem[Hines(2018)]{hines2018stepping}
J.~Hines.
\newblock Stepping up to summit.
\newblock \emph{Comput. Sci. Eng.}, 20\penalty0 (2):\penalty0 78--82, 2018.

\bibitem[Borrell et~al.(2013)Borrell, Sillero, and
  Jim{\'e}nez]{borrell2013code}
G.~Borrell, J.~A. Sillero, and J.~Jim{\'e}nez.
\newblock A code for direct numerical simulation of turbulent boundary layers
  at high {R}eynolds numbers in {BG/P} supercomputers.
\newblock \emph{Comput. Fluids}, 80:\penalty0 37--43, 2013.

\bibitem[Lee et~al.(2013)Lee, Malaya, and Moser]{lee2013petascale}
M.~Lee, N.~Malaya, and R.~D. Moser.
\newblock Petascale direct numerical simulation of turbulent channel flow on up
  to 786k cores.
\newblock In \emph{Proc. ACM/IEEE Supercomputing Conf.}, pages 1--11. IEEE,
  2013.

\bibitem[CSCS(2019)]{website:cscs}
CSCS.
\newblock \url{https://www.cscs.ch/}, 2019.

\end{thebibliography}

\end{document}